\documentclass{article}

\usepackage{arxiv}

\usepackage[utf8]{inputenc} 
\usepackage[T1]{fontenc}    
\usepackage{hyperref}       
\usepackage{url}            
\usepackage{booktabs}       
\usepackage{amsfonts}       
\usepackage{nicefrac}       
\usepackage{microtype}      
\usepackage{lipsum}		
\usepackage{graphicx}
\usepackage{natbib}
\usepackage{doi}
\usepackage{commands}

\title{Variance as a predictor of health outcomes: Subject-level trajectories and variability of sex hormones to predict body fat changes in peri- and post-menopausal women}

\author{ \href{https://orcid.org/0000-0002-9366-8506}{\includegraphics[scale=0.06]{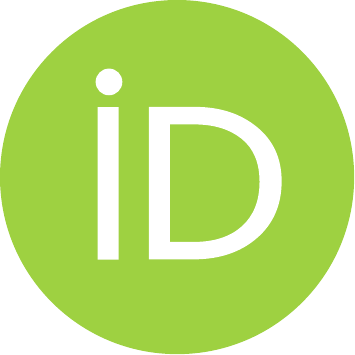}\hspace{1mm}Irena Chen}\\
	Department of Biostatistics\\
	University of Michigan\\
	\texttt{irena@umich.edu} \\
	\And
	\href{https://orcid.org/0000-0000-0000-0000}{\includegraphics[scale=0.06]{orcid.pdf}\hspace{1mm}Zhenke Wu} \\
	Department of Biostatistics\\
	University of Michigan\\
	\And
	Siobàn D. Harlow\\
	Department of Epidemiology \\
	University of Michigan\\
	\And
	Carrie A. Karvonen-Gutierrez \\
	Department of Epidemiology \\
	University of Michigan\\
		\And
	Michelle M. Hood \\
	Department of Epidemiology \\
	University of Michigan\\
	\AND
    Michael R. Elliott \\
	Department of Biostatistics\\
	University of Michigan\\
}

\renewcommand{\shorttitle}{Variance as a predictor}

\hypersetup{
pdftitle={Variance as a predictor of health outcomes},
pdfsubject={Methodology, applications},
pdfauthor={Irena Chen, Zhenke Wu,Siobàn D. Harlow, Carrie A. Karvonen-Gutierrez, Michelle M. Hood, Michael R. Elliott},
pdfkeywords={Estradiol, Follicle-stimulating hormone , Hamiltonian Monte Carlo, Joint models, Menopause, Subject-level variability,  Study of Women's Health Across the Nation (SWAN), Variance component priors},
}

\begin{document}
\maketitle

\begin{abstract}
	Longitudinal biomarker data and cross-sectional outcomes are routinely collected in modern epidemiology studies, often with the goal of informing tailored early intervention decisions. For example, hormones such as estradiol (E2) and follicle-stimulating hormone (FSH) may predict changes in womens' health during the midlife. Most existing methods focus on constructing predictors from mean marker trajectories. However, subject-level biomarker variability may also provide critical information about disease risks and health outcomes. In this paper, we develop a joint model that estimates subject-level means and variances of longitudinal biomarkers to predict a cross-sectional health outcome. Simulations demonstrate excellent recovery of true model parameters. The proposed method provides less biased and more efficient estimates, relative to alternative approaches that either ignore subject-level differences in variances or perform two-stage estimation where estimated marker variances are treated as observed. Analyses of women’s health data reveal larger variability of E2 or larger variability of FSH were associated with higher levels of fat mass change and higher levels of lean mass change across the menopausal transition.
\end{abstract}


\section{Introduction}\label{sec:intro}

 The menopausal transition is a critical lifestage that can shape women's midlife and long-term health. Reproductive aging and the menopausal transition are characterized by decreasing levels of estradiol (E2) and increasing levels of follicle-stimulation hormone (FSH) \citep{randolph_change_2004}.  These two hormones, in addition to having important roles in regulating reproductive functionality, are also important for regulating other health outcomes \citep{karvonen-gutierrez_menopause_2017}. E2 is known to regulate fat mass and a decline in E2 is associated with increased fat mass \citep{colleluori_fat_2018}. 
Further, across the menopausal transition, women with obesity had lower E2 levels during the premenopause, but higher levels of E2 during post-menopause as compared to women without obesity \citep{randolph_change_2011}. FSH has also been linked to health outcomes in women; for example, higher levels of FSH may be associated with increased risk of osteoporosis \citep{zaidi_fsh_2018}.  Additionally, FSH appears to be an important predictor of increased adiposity, reduced energy expenditure \citep{sponton_burning_2017, kohrt_preventing_2017} and lower lean mass during the postmenopause \citep{gourlay_follicle-stimulating_2012}.  Data from animal models suggest that blockage of the FSH receptor in mice was sufficient to reduce body fat and increase lean mass. \citep{liu_blocking_2017}.

\par A large and well-developed body of literature over the past two decades uses longitudinal biomarker or questionnaire data to predict health outcomes: an early example is given by \citet{henderson_joint_2000}, who tied psychiatric disorder measures over time to prediction dropout in schizophrenia trials via a subject level random effect in the disorder trajectory that is also present as a frailty in the time-to-event model. \citet{proust-lima_joint_2014} link latent classes of prostate specific antigen to survival models. More recent work by \citet{wang_dynamic_2017} considers multiple longitudinal predictors - in their case, measures of daily functioning - to predict onset of Parkinson's disorder. However, methods that assess the ability of measures of residual variability are largely lacking in the literature, despite calls for increased focus on developing methods for variance structures. Such methods may elucidate whether systemic dependence of variability on known factors could yield both better prediction and improved inference \citep{carroll_variances_2003}. This oversight appears to be a substantial gap in the statistical methods literature given that increased variance can be an early predictor of instability in biological systems; for example, heart rate variability may be a marker of autonomic dysregulation given its predictive nature with poor health outcomes \citep{young_heart-rate_2018}. 

The existing literature often uses a simple two-stage approach with a squared-error estimate of variance obtained from observed data to predict a single outcome, ignoring the inherent uncertainty in the constructed variance estimates. Joint modeling efforts are rarer but critical given that statistical uncertainty of the constructed predictors, e.g., mean and residual variance estimates, can lead to extremely biased estimates of association and uncontrolled sampling variances  \cite[e.g.,][]{ogburn2021warning, wang2020methods}. Finally, prediction approaches based on multiple trajectories - which allow for consideration of residual covariances as well as residual variances as predictors -- appears to be completely absent from the biostatistics toolkit. 

\par This paper is organized as follows. Section \ref{sec::prev_variances} provides a brief overview of previous work on variance as a predictor of health outcomes, highlighting the statistical consequences of ignoring the statistical uncertainty in multiple predictors for the outcome. Section \ref{sec::model_form} describes our joint model framework linking the mean and variance trajectory model for the longitudinal predictors with the model for the outcome. 
Section \ref{sec:Simulation} conducts extensive simulation studies to show that our proposed approach produces less biased estimates of the outcome regression coefficients with proper control of the sampling variances relative to a variety of increasingly complex two-stage competitors that fail to account for the statistical uncertainty 
in the subject-level means and in the variances. Section \ref{sec::swan_application} applies our method to assess the associations between the mean trajectories, variance, and covariance of E2 and FSH and changes in fat and lean body mass during the menopausal transition using data from the Study of Women's Health Across the Nation. Section \ref{sec::discussion} discusses the implications of our work along with limitations and directions for future research. 

\section{Previous Work on Variances as a Predictor of Outcomes}
\label{sec::prev_variances}

Joint modeling of longitudinal trajectories and cross-sectional outcomes is a rich area of statistical research \citep{chi_joint_2006, ibrahim_basic_2010, lawrence_gould_joint_2015}. For example, including longitudinal measurements in joint longitudinal - survival models can provide better estimates of the survival outcome by accounting for the individual trajectories over time, as shown by \cite{long_joint_2018}, who used a joint model to predict motor diagnosis using longitudinal characterizations of Huntington's disease progression taken at annual visits; also see \citet{Papageorgiou2019} for a comprehensive review. Until recently, most of the focus in the area of modeling individual trajectories has been on modeling the mean trends. The variability associated with individual biomarkers has traditionally been treated as a nuisance parameter. 

However, interest in both modeling the residual variability of individual trajectories and using these estimates as predictive variables has been growing. \citet{elliott_associations_2012} studied the relationship between individual variability in short-term memory tests and long term onset of senility. They found that increased variability in the memory tests were associated with increased risk of senility. Furthermore, this variability was a stronger predictor of senility than the mean trajectories. With regards to women's health, some previous research has considered variance as a predictor of health outcomes. \citet{harlow_analysis_2000} found that women who had increased menstrual cycle variability at a younger age were more likely to experience abnormal uterine bleeding. The variability of menstrual cycle length was also found to be an important predictor of menopausal onset \citep{huang_modeling_2014}. Furthermore, \citet{sammel_models_2001} found that E2 variability during the menopausal transition was highly predictive of experiencing hot flashes. \citet{jiang_joint_2015} modeled the individual mean and variances of FSH hormone levels over time to predict risk of hot flash and found that the variance of FSH was strongly associated with the risk.  

Our work makes two primary contributions to the existing statistical literature: 1) we model individual-level variances and co-variances of multiple longitudinal trajectories by proposing flexible hierarchical priors for residual variance-covariance matrices that may flexibly vary by subject and 2) we incorporate these subject-level means and variance-covariance matrices for longitudinal data and a cross-sectional outcome in a joint modeling framework, rather than a two-stage approach that we show to be statistically inferior. Also, to the best of our knowledge, this is the first rigorous assessment of the role of individual variability of E2 and FSH hormones in predicting body mass composition changes.  

\subsection{Measurement Error in Multivariate Linear Models}
\label{sec::measurement_error}
An advantage of joint models relative to two-stage models is that the uncertainty associated with the parameters estimated in the first stage is carried over to the second stage. 
Consider a simple linear relationship of the form: 
\begin{equation}
        Y = X\beta  + \epsilon,
\label{eq::multiple_reg}
\end{equation} 
where $X$ is a $n \times K$ matrix of $K$ predictors and $\epsilon$ is an $n \times 1$ vector of independent normal error terms with mean 0 and variance $\sigma^2$. Suppose that the true relationship between $Y$ and $X$ is described by (\ref{eq::multiple_reg}), but instead, we observe $\Tilde{X}$, where $\Tilde{X} = X + U$, where $U$ is the matrix of normally-distributed independent measurement errors with mean 0 and variance-covariance $\Sigma_U$.

If $U \indep X$, then in the $K=1$ scenario, we know that the estimate of $\beta$ will be attenuated towards the null \citep[~p.42-43]{carroll_measurement_2006}. For $K > 1$, with multiple predictors measured with error, the estimates of the $\beta$ are still biased, but the direction of the bias now depends on the correlation between the measurement errors \citep[~p.53-55]{carroll_measurement_2006}. 
Consider the following equation for $K=2$ predictors:

\begin{equation}
        Y = \alpha + \beta_1 X_1 + \beta_2 X_2  + \epsilon,
\label{eq::two_predictors}
\end{equation}
and suppose we measure $X_1, X_2$ with some error:
\begin{equation}
         \Tilde{X}_1 = X_1  + U_1,\Tilde{X}_1 = X_2  + U_2,
\label{eq::two_predictors_error}
\end{equation}
\citet[~p.1477–1479]{griliches_chapter_1986} derive the bias of estimating $\beta_1$ as: 
\begin{equation}
    \textrm{plim}(\hat{b}_{1} - \beta_1) = -\dfrac{\beta_1\lambda_1}{(1-\rho^2)} + \dfrac{\beta_2\lambda_2\rho}{(1-\rho^2)},
\end{equation}
where $\hat{b}_1$ is the coefficient obtained from regressing $Y$ on $\Tilde{X}_1$ in the multiple regression model, $\lambda_1 =  \mathrm{Var}(U_1)/\sigma^2_{1}$ is the relative amount of measurement error in $X_1$ where $\sigma^2_{1}$ refers to the variance of $X_1$, and $\rho$ is the (true) correlation between $X_1, X_2$; $plim$ refers to convergence in probability. A similar equation can be derived for the bias of estimating $\beta_2$ in the presence of such measurement errors. We can see, then, that the bias is increased by a factor of $\dfrac{1}{(1-\rho^2)}$. The overall effect of the additional variable $\Tilde{X}_2$ is a  bias towards the null \cite[~p.1479]{griliches_chapter_1986}. For $K >2$ variables, the expressions for the bias of each predictor become more complicated to derive. We show via simulations in Section \ref{subsec::other_methods} that the bias in the mean parameters is clear in the two-stage model linear regression alternatives to the joint model. 
Furthermore, we see that that bias in the estimates of the variance-covariance parameters persists in the two-stage model alternatives. This would be an issue if we believe that individual variability (covariability) is actually predictive of an outcome of interest. 

\section{Proposed Model}
\label{sec::model_form}
\subsection{Notation} Let $\cD = (Y_i,\bX_{ij},t_{ij}, \bW_i)$ be the observed data for subject $i = 1, \ldots, N$, where $Y_i\in \RR^1$ is a continuous outcome, $\bX_{ij} = (X_{ij1}, \ldots, X_{ijQ})^\transp$ is a vector of $Q$ time-varying marker values at observation time points $t_{ij}, j=1, \ldots, n_{i}$, that may differ by subjects, and $\bW_i = (W_{i1}, \ldots, W_{id})^{\transp}$ is a vector of $d$ time-invariant covariates, e.g., race/ethnicity, activity class.

The proposed model has two connected components. We first specify the regression model for irregularly and longitudinally observed multiple markers $(t_{ij}, \bX_{ij}),j=1, \ldots, n_{ij}$; the second component links the outcome $Y_i$ to time invariant covariates $\bW_i$ and unobserved individual-specific vectors of regression coefficients and residual variance-covariance matrices in the model for longitudinal markers, enabling inference about how the mean trajectories and residual variations are associated with the outcome $Y_i$. 
\subsection{Likelihood}
\label{sec::model_likelihood}

\subsubsection{Component 1: Longitudinal Markers} We specify the model for the longitudinal marker data as follows:
\begin{align}
  \quad \bX_{ij} \mid \Bb_i, \Sb_i & {~\sim~}  \mathcal{N}_Q\left(\bmu(t_{ij}; \Bb_i), \Sb_i\right), j=1, \ldots, n_{i}, {\sf ~independently~for~} i=1, \ldots, N, \label{eq::marker_model}\\
  \bb_{iq}  & \overset{\sf indep.}{\sim}  \mathcal{N}_P(\bbeta_q, \Sigma_q), q=1, \ldots, Q 
  \label{eq::individual_beta_sigma}
\end{align}
where $\cN_Q(\bmu,\Sb)$ is a generic notation that represents a $Q$-dimensional multivariate Gaussian distribution with mean vector $\bmu$ and variance-covariance matrix $\Sb$; $\bmu(t; \Bb_i)$ is a $Q$-dimensional function of time given by $\Bb_i = [\bb_{i1},...,\bb_{iQ}]^\transp$ and $\bb_{iq}=(b_{iq1}, \ldots, b_{iqP})^{\transp}$ is a vector of $P$ regression coefficients for the $q$-th marker.  Here $P$ is the number of basis functions of time; to simplify presentation in this paper, we assume the same number of coefficients for each marker, e.g., intercept and slope. In addition, $\bbeta_q = (\beta_{q1}, \ldots, \beta_{qP})^\transp$ is a vector of population mean regression coefficients that are specific to the $q$-th marker. In addition, Equation (\ref{eq::individual_beta_sigma}) has assumed individual-specific regression coefficients for any two markers are conditionally independent.

\subsubsection{Component 2: Outcome Regression (OR) Model}

The outcome variable $Y_i$ is assumed to be related to individual-specific  mean and variance-covariance parameters $\Bb_i$ and $\Sb_i$ in the longitudinal marker model (\ref{eq::marker_model}) as follows:
\begin{align}
  Y_i \mid \Bb_i, \Sb_i, \bW_i\sim &  \mathcal{N}\left(\eta_i, \sigma^2\right), i=1, \ldots, N, \label{eq::outcome_regression}
\end{align}
where $\bW_i=(W_{i1},\ldots, W_{id})^\transp$ is a vector of $d$ time-invariant covariates; $\eta_i = \eta(\Bb_i,\Sb_i, \bW_i; \balpha, \bgamma,\bgamma^W)$ and $\eta(\cdot; \balpha, \bgamma, \bgamma^W)$ is a generic mean outcome regression function parameterized by $\balpha$ (for $\Bb_i$), $\bgamma$ (for $\Sb_i$), and $\bgamma^W$ (for $\bW_i$); $\balpha$, $\bgamma$, and $\bgamma^W$ are of dimension $PQ$, $Q(Q+1)/2$, $d$, respectively. In this paper, we will illustrate the statistical performance of such a formulation by focusing on simple specifications of $\eta(\cdot)$, e.g., linear models. The framework readily generalizes to general outcome models; in Section \ref{sec::mixture_model_swan}, we illustrate a model with a Gaussian scale-mixture outcome regression model for lean mass rate-of-change outcomes.

\subsection{Priors}
\label{subsec::priors}

In this section, we specify the prior and hyperprior distributions for the unknown parameters in the two likelihood components.

\subsubsection*{Prior for $\Sb_i$} We rewrite $\Sb_i = \Db_i \Rb_i \Db_i$,
where $\Db_i=\diag(d_{i1},...,d_{iQ})$ is a diagonal matrix of residual variances and $\Rb_i$ is the associated subject-level correlation matrix. We assume
\begin{align}
   \log(d_{iq})  &~ {\sim} ~\mathcal{N}(\nu_q, \psi_q^2),  \label{eq::di_hyperprior}
\end{align}
independently for marker $q=1, \ldots, Q$, and subject $i=1, \ldots, N$. Because $\Rb_i$ is a correlation matrix, we only need to specify the prior distribution for the off-diagonal elements. We consider the special case of $Q=2$, where $r_{12i}$ is unconstrained, and separately the general case of $Q>2$, where the components of $\Rb_i$ must meet the positive definite criterion.
 
\subsubsection*{Special Case: $Q=2$.}
For a $2\times2$ correlation matrix, we place the following prior on the off-diagonal value $r_{12}$: 
\begin{align}
   (r_{i12} +1)/2  & \sim {\sf Beta}(a'_{12}, b'_{12}), {\sf ~independently~for~} i=1, \ldots, N.
   \label{beta_hyperprior}
\end{align}

Finally, we specify hyperpriors for $\nu_q$, $\psi_q$ and $(a'_{12},b'_{12})$ by
\vspace{0cm}
 \begin{align}
     \nu_q  & \sim \cN(m, \xi_q^2), \psi_q \sim \textsf{half-Cauchy}(0,\tau),{\sf~independently~for~}q\leq Q,\\ 
    a'_{12} & \sim {\sf Exp}(\kappa), ~ b'_{12} \sim {\sf Exp}(\kappa').
    \label{hyperprior_QiRi}
    \end{align} 

In our application, we set $m=0$ and $\xi_q=10$.  In addition, we use a half-Cauchy hyperprior on $\psi_q$ instead of the inverse-Gamma distribution as this prior is recommended for datasets where the variance $\psi_q$ may be small \citep{gelman_prior_2006}. In this setting, inferences using the inverse-Gamma distribution are extremely sensitive to the choice of hyperparameter values \citep[p.~524]{gelman_prior_2006}, which makes the inverse-Gamma prior ``not at all uninformative''.  
The half-Cauchy distribution avoids this potential issue due to its heavier tail, which still allows for higher variance values.  

\begin{remark}
The defining feature of our framework is the individual-specific variance-covariance matrices, $\{\bS_i,i=1,\ldots, N\}$, over which we must specify a hierarchical prior distribution. Such priors must not be restrictive in capturing between-subject differences and similarities in the variance-covariance matrices. Focusing on the prior specification for individual-specific correlation matrices $\{\Rb_i,i=1,\ldots, N\}$, standard priors designed for a single unknown population correlation matrix, e.g., Lewandowski-Kurowicka-Joe (LKJ) prior \citep{lewandowski2009generating}, have severe drawbacks. In particular, the LKJ distribution is governed by a single positive scale parameter, $\zeta$, that tunes the strengths of the correlations. The off-diagonal elements of a $K\times K$ correlation matrix are marginally distributed as: $(r_{ilk}+1)/2 \sim \sf Beta (a,b)$, where $a = b = \zeta -1 +K/2$. This implies that the correlations will \textit{a priori} be concentrated around $0$. However, in our motivating application, $\{r_{ilk},i=1,\ldots, N\}$ represent the individual-specific residual correlations between the $l$- and the $k$-th hormone, which 1) by domain knowledge are a priori unlikely to have a strong prior of being near zero, and 2) may vary between subjects in a way far from the implied distribution of $2\cdot {\sf Beta} (a,b)-1$. In Equation (\ref{hyperprior_QiRi}), we have removed this identity restriction and specified hyperpriors on $a$ and $b$, which provides greater flexibility in allowing the data to estimate the true $a$ and $b$. The same argument can be applied to the Inverse-Wishart distribution for variance-covariance matrices, which is also governed by a single scale parameter.
\end{remark}

In addition, we argue that the Exponential distribution in Equation (\ref{hyperprior_QiRi}) is 
a sensible hyperprior on the Beta parameters as follows: Let $x_{1}, \ldots, x_{n}$ be data from 
a ${\sf Beta}(a, b)$ distribution. Assume that $a \sim {\sf Exp}(\lambda_a), b \sim {\sf Exp}(\lambda_b)$. 
We can also assume $a$ and $b$ are independent \textit{a priori}. Then the posterior distribution, 
$p(a, b | x) \propto L(a, b) p(a) p(b) \propto \prod^{n}_{i=1} \exp(-\ln B(a, b) + 
  (a -1)\ln(x_i) +  (b -1)\ln(1-x_i) \times \exp(\ln(\lambda_a) - \lambda_a a + \ln(\lambda_b)- \lambda_b b) 
\propto \exp(-n\ln B(a, b) + a \sum^{n}_{i=1} (\ln x_{i} - \lambda_a) + b \sum^{n}_{i=1} (\ln(1- x_{i}) - \lambda_b)$, 
which suggests that the posterior distribution of $a, b$ would be updated from a flat prior by subtracting $(\lambda_a, \lambda_b)$ from the sufficient statistics $ \sum^{n}_{i=1} \ln x_{i} $ and $\sum^{n}_{i=1} \ln(1- x_{i})$. 
The reason for this is that the Maximum Likelihood Estimators (MLEs) for $a, b$ respectively can be approximated by: 
  $\hat{a}_{MLE} = \dfrac{1}{2} + \dfrac{\hat{G}_{x}}{1-\hat{G}_{x} - \hat{G}_{(1-x)}}, 
\hat{b}_{MLE} = \dfrac{1}{2} + \dfrac{\hat{G}_{(1-x)}}{1-\hat{G}_{x} - \hat{G}_{(1-x)}}$ where $\hat{G}_{x} = \exp(n^{-1}\sum^{n}_{i=1} \ln(x_{i})),
\hat{G}_{1-x} = \exp(n^{-1}\sum^{n}_{i=1}\ln(1-x_{i}))$, where $(a, b) > 1$. The posterior modes using the Exponential priors become $\dfrac{1}{2} + 
  \dfrac{\hat{G}_{x}\exp(\lambda_{a})}{1-\hat{G}_{x}\exp(\lambda_{a}) - \hat{G}_{(1-x)}\exp(\lambda_{a})}, \dfrac{1}{2} + \dfrac{\hat{G}_{(1-x)}\exp(\lambda_{b})}{1-\hat{G}_{x}
    \exp(\lambda_{b}) - \hat{G}_{(1-x)}\exp(\lambda_{b})} $.  When $\lambda_{a}, \lambda_{b} \xrightarrow[]{} \infty  $, the posterior modes of $a, b$ shrink towards $\dfrac{1}{2}$, which is the Jeffrey's prior. 
When $\lambda_{a}, \lambda_{b} \xrightarrow[]{} 0 $,
we recover the likelihood. Overall, these results suggest that the choice of the Exponential distribution is a flexible hyperprior on $a, b$ and thus is a reasonable choice. 

\subsubsection*{General Case: $Q\geq 3$.} 
\label{sec::3biomarker}

For the general case of a $ Q \times Q$ correlation matrix where $Q \geq 3$, the off-diagonal values of the individual correlation matrices are now more complicated to estimate since the space of valid correlation matrices is a proper subset of the space of all possible $ Q \times Q$ matrices. We address this constraint by following the approach of \cite{ghosh_bayesian_2021} where the off-diagonal values are parameterized in terms of hyperspherical coordinates. The angles are allowed to vary freely over $[0,\pi]$ before being back-transformed into valid correlation values. To illustrate, we specify the prior for $\Rb_i$ for when $Q=3$ as follows (similarly for $Q>3$): 
\begin{equation*}
 r_{12} = \cos(\theta_{12}),r_{13} = \cos(\theta_{13}),r_{23} = \sin(\theta_{12})\cdot \sin(\theta_{13})\cdot \cos(\theta_{23})+\cos(\theta_{12})\cdot \cos(\theta_{13}),\label{eq::3biomarker_corr_values2}
\end{equation*}
where $\theta_{12}  = \arccos(c_{12}),  \theta_{13} = \arccos(c_{13}), \theta_{23} = \arccos(c_{23})$, and 
\begin{eqnarray*} 
(c_{i12} +1)/2&\sim {\sf Beta}(a'_{12}, b'_{12}), (c_{i13} +1)/2 \sim {\sf Beta}(a'_{13}, b'_{13}), (c_{i23} +1)/2 \sim {\sf Beta}(a'_{23}, b'_{23}). \label{eq::3biomarker_beta}
\end{eqnarray*}


As in the case where $Q =2$, we specify hyper-priors for $a'_{kl}, b'_{kl}, k < l $, e.g., the Exponential prior. 

\paragraph{Priors for population longitudinal marker regression coefficients:}

\begin{align}
\label{prior_Bis}
   \bbeta_{q} & ~{\sim}~ \mathcal{N}_P(0, \xi_q^2I_{P\times P}), {\sf ~independently~for~} q=1, \ldots, Q \\\label{eq::beta_priors}
   \Sigma_q   & = \Kb_q \Lb_q \Kb_q, ~\Kb_q  =\diag\{k_{q1}, \ldots, k_{qP}\},q = 1, \ldots, Q,\\
    k_{qp}    & \sim \textsf{half-Cauchy}(0, \tau_0), p = 1, \ldots, P, {\sf~and~} \Lb_q \sim {\sf LKJ}(\zeta), {\sf ~independently~for~} q=1, \ldots, Q\label{eq::KiLi_priors}
\end{align} where $\Kb_q=\diag\{k_{q1}, \ldots, k_{qP}\}$ is a diagonal matrix and $\Lb_q$ is a correlation matrix. The $\tau_0, \zeta$ parameters are set in practice as 2.5 and 1 respectively. It is fine to use the half-Cauchy and the LKJ priors in Equation (\ref{eq::KiLi_priors}) since, for each marker $q$, they are standard hyperpriors for a  \textit{single} population variance matrix $\Kb_q$ and a \textit{single} population-level correlation matrix $\Lb_q$, which is different from Equation (\ref{hyperprior_QiRi}) that specifies the prior for multiple and individual-specific variance-covariance matrices. In addition, \citet{gelman_prior_2006} shows that the Inverse-Gamma prior distribution is overly informative for a hierarchical variance and this issue persists for an Inverse-Wishart on a hierarchical covariance matrix. We therefore choose to specify half-Cauchy for variances $\{k_{qp},p=1,\ldots, P\}$, and the LKJ distribution for $\Lb_q$.
\paragraph{Prior for parameters in the outcome regression model}
\label{prior_outcome_model}
\par For the outcome model, we place diffuse  independent Gaussian priors for each element of the outcome regression parameters ($\balpha$, $\bgamma$, $\bgamma^W)$. Finally, to complete the prior specification, we place a diffuse prior on the outcome residual standard deviation parameter $\sigma \sim \textsf{half-Cauchy}(0,\tau_1)$. In our simulation studies and data analysis, we set the priors on the regression parameters as $\mathcal{N}(0, 10^2)$ (a very diffuse prior in order to allow the data to estimate the parameters) and $\tau_1 = 2.5$ (the default suggested by \citet{carpenter_stan_2017}).  

Let $\Theta = (\bbeta_q, \Sigma_q,\xi, \nu_q, \psi_q, a'_{kl}, b'_{kl}, \balpha, \bgamma,\bgamma^W, \sigma) $ denote the unknowns of interest in the proposed model. Let $\pi(\Theta)$ denote the prior distribution where we have assumed that all parameters in $\Theta$ have independent components: 

$\pi(\Theta)=\Pi^{Q}_{q=1} [\pi(\bbeta_q) \pi(\Sigma_q) \pi(\xi_q)  \pi(\nu_q, \psi_q)] \Pi_{k<l} [\pi( a'_{kl}, b'_{kl})] \pi(\balpha, \bgamma,\bgamma^W)) \pi(\sigma) $.

\paragraph{Joint Distribution}  The joint distribution of the data and unknowns is
\begin{align}
   P(\Theta, \cD)  \propto &  \prod^{n}_{i=1}   \prod^{Q}_{q=1} \Big\{ \frac{1}{\sqrt{(2\pi)^{q}|\Sigma|}}{\exp}(-\frac{1}{2}(\bb_{iq}-\bbeta_q)^\transp\Sigma^{-1}(\bb_{iq}-\bbeta_q)) \nonumber
    \\ 
    & \times \frac{1}{\sqrt{2\pi\xi_q^{2}}}{\exp}\left[\frac{(\log(d_{iq})-\nu_q)^{2}}{2\xi_q^{2}}   \right]  \left( \frac{[(r_{ikl}+1)/2]^{a'_{kl}-1}\{1-([(r_{ikl}+1)/2]\}^{b'_{kl}-1}}{{\sf Beta}(a'_{kl},b'_{kl})} \right) \nonumber
     \\
     & \times \prod ^{n_i}_{j=1} \frac{1}{\sqrt{(2\pi)^{n_i}|\Sb_i|}}{\exp}\left(-\frac{1}{2}\{\bX_{ij}-\bmu( t_{ij}; \Bb_i)\}^\transp \Sb_i^{-1}\{\bX_{ij}-\bmu(t_{ij}; \Bb_i)\}\right) \Big\} \nonumber 
     \\
    & \times \frac{1}{\sqrt{2\pi\sigma^{2}}}{\exp}\left[\frac{(Y_{i}-\eta(\Bb_i, \Sb_i, \bW_i; \balpha,\bgamma,\bgamma^W))^2}{2\sigma^{2}} \right] \times \pi(\Theta). \label{posterior_likelihood}
\end{align}

Figure S1 in the Supplementary Materials \citep{chen_supp_2022} uses a directed acyclic graph to visualize and summarize the hierarchical relationships between the different components of our modeling framework.

\subsection{Posterior Inference}

In a Bayesian framework, the inference is conducted based on the posterior distribution $P(\Theta \mid \cD)$. However, it is not feasible to derive the closed-form posterior distribution owing to the lack of prior-likelihood conjugacy in our proposed model. We therefore used Hamiltonian Monte Carlo to draw sequential samples and approximate the posterior distribution. We implement the model using Stan and the \verb"rstan" package \citep{rstan} as the interface for running the model and obtaining the posterior estimates. Code to run the joint model and generate the data used in our simulation studies are provided in attached supplementary files. For our simulations studies in Section \ref{sec:TwoBiomarkersSim}, \ref{sec:ThreeBiomarkersSim} and Section 3 in the Supplementary Material \citep{chen_supp_2022}, we run two chains per independent replicate data set, with $2,000$ iterations and $1,000$ burn-in. 
For the data application in Section \ref{sec::swan_application}, we ran 4 chains each and 4,000 iterations with 2,000 burn-in. Visual inspection of the traceplots for all model parameters indicated non-divergent chains. All chains were combined for calculating posterior summaries.

We also examined Stan's R-hat convergence diagnostic \citep{vehtari_rank-normalization_2021} and the Effective Sample Sizes (ESS) to determine if the chains had mixed well. The R-hat value for all model parameters was less than 1.05. In the fat mass rate of change model, the ESS for all of the model parameters was at least 100 times the number of chains used, except for the model parameter corresponding to individual-level E2 variance and the parameter corresponding to fat mass proportion at the first visit. This was also the case for the lean mass rate of change model (E2 variance and lean mass proportion at the first visit).  However, the ESS values for these parameter were still above the recommended sample size number, $5\hat{m}$, where $\hat{m}$ is the number of chains  \citep[~p.384]{gelman_bayesian_2013}. The R-hat values of these parameters was also effectively 1.00 in both models. Based on these diagnostics, we concluded that our models had converged. The posterior predictive checks we conducted (see S@ in the Supplementary Material, \cite{chen_supp_2022}) suggest that our model generates reasonable estimates for the observed outcomes and trajectories.

\section{Simulation Study}
\label{sec:Simulation}
We conducted simulation studies to 1) evaluate our model's operating characteristics and 2) compare against common alternatives that could also be used in modeling individual means and variances as predictors of outcomes.
We evaluated our model performance across independent simulation replicates using three criteria: for each method and each parameter $\theta$, we assess the 1) bias (defined as $1/R\sum_{r=1}^{R}(\hat{\theta}^{(r)} - \theta_0)$ where $\hat{\theta}^{(r)}$ is the posterior mean of $\theta$ obtained from the $r$-th replication), 2) the coverage rate of the nominal $95\%$ credible intervals (CrI; defined as $1/R \sum^{R}_{r=1} \mathbbm{1}\{\theta_0 \in I_r\} $ where 
$I_r$ is the $95\%$ CrI for parameter $\theta$ obtained by computing the 2.5\% and 97.5\% percentiles of the draws from the posterior distribution for the $r$-th replication, and 3) average length of the 95\% CrIs obtained across simulation replicates, defined as $1/R \sum^{R}_{r=1} T_{r} $, where $T_{r}$ is the length of $I_r$, i.e., the range of the estimated 2.5\% and 97.5\% posterior quantiles for $\theta$ in replicate $r$.

In this section, we present the results from simulation studies with $Q=2$ and $Q=3$ biomarkers. Additionally, in Section S3 in the Supplementary Material \citep{chen_supp_2022}, we present a simulation study examining our model performance when we approximate the true nonlinear relationship between the marker means and variances and the outcome as linear. 

\subsection{Simulation 1: Two Biomarkers.}
\label{sec:TwoBiomarkersSim}
In this simulation, we assume the mean  trajectories can be expressed linearly with individual intercepts and slopes. We generated $n_i=6$ to $15$ time points for $N=1,000$ individuals, which mimics the data used in Section \ref{sec::swan_application}. We then simulated two trajectories for each individual using the following parameters.

\subsubsection*{Component 1: Longitudinal Markers }
\begin{align*}
& X_{itq}  = b_{iq1} + b_{iq2} t + \epsilon_{iq}, q=1,2; \bB_{i1} \sim \mathcal{N}_2\left(\bbeta_{1}, \Sigma_{1}\right),
\bB_{i2} \sim \mathcal{N}_2\left(\bbeta_{2}, \Sigma_{2}\right)\\
& \bbeta_{1} =(0, 2)^\transp, 
\bbeta_{2} = (2, 1)^\transp,
\Sigma_1 =\begin{pmatrix}
1 & -0.05 \\
-0.05 & 1 
\end{pmatrix},
\Sigma_2 = \begin{pmatrix}
1 & -0.1 \\
-0.1 & 0.5 
\end{pmatrix}, \\ 
& \log(d_{i1}) \sim \mathcal{N}(0, 0.75)/2,  \log(d_{i2}) \sim \mathcal{N}(0.5, 0.5)/2,
(r_{i12} +1)/2  \sim {\sf Beta}(1,5). \label{sim1:marker_generating}
\end{align*}

\subsubsection*{Component 2: Outcome Regression Model} To generate the outcome for each individual, we assume $Y_i \sim \cN(\eta(\Bb_i, \Sb_i), \sigma^2)$ and set 
\begin{align*}
  & \eta(\Bb_i, \Sb_i) = \alpha_{11} b_{i11}  +\alpha_{12} b_{i12} +  \alpha_{21} b_{i21} + \alpha_{22} b_{i22} + \gamma_{11} s_{i11} + \gamma_{21} s_{i21} +  \gamma_{22} s_{i22},
\end{align*}
where the true values of $\balpha, \bgamma$ are  $\balpha = (-3,-3,-3, 3), \bgamma = (2,-1,2)$; we did not include other time-invariant covariates $\bW_i$. These particular truth values were chosen so that the distribution of the outcome $Y_i$ would be similar to the distribution of the SWAN body mass outcomes (our data analysis application). Lastly, we set $\sigma^2 = 1$. We present the results for $R=200$ replicates in Table \ref{tab:simulation_comparisons_1000_ids} for the outcome submodel parameters $\balpha$, $\bgamma$. See Table S5 in the Supplementary Materials \citep{chen_supp_2022} for the results for the other model parameters. 

\subsubsection{Alternative Methods}
\label{sec:alternatives}
We briefly introduce three common alternatives in our comparative simulation study: two-stage simple linear model (TSLM), two-stage linear mixed model (TSLMM), and two-stage individual-variance (TSIV) model. We refer to our joint model as the ``Joint Model with Individual Variances", or JMIV. 

\label{subsec::other_methods}
\paragraph{Two-Stage Simple Linear Regression (TSLM)} One of the most simple alternative models we could use is the linear regression model in two stages. We used the \verb"lm()" function in R and first fit the following model: 
\begin{align*}
{X_{itq}}  = \beta_{iq0} + \beta_{iq1} t + \epsilon_{iq}, q = 1,2.
\end{align*} 
Here, we will obtain $\hat{\beta}_{iq0}, \hat{\beta}_{iq1}$ via ordinary least squares estimates for the mean parameters $b_{iq0},b_{iq1}$. 
To estimate $s_{i11}, s_{i22}$, we collected the residuals from each regression, e.g.,  $ r_{ij1} =  (x_{it1} - ( \hat{\beta}_{i10}+ \hat{\beta}_{i11} t_{ij})), j= 1,\ldots, n_i$, and computed the sample variance of these residuals, which we term ``estimated residual variance" for each individual, i.e., $\hat{s}_{i11}$; similarly, we obtain $r_{ij2}$ and $\hat{s}_{i22}$. The residual covariance, $\hat{s}_{i12}$, was estimated by sample covariance of $(r_{ij1}, r_{ij2})$.
We then used linear regression to model the outcome based on the estimated coefficients and residual variances and covariances from the first-stage model: $\mathop{{}\mathbb{E}}(Y_{i}\mid {\sf others})  = \sum_{q=1,2}\left\{\alpha_{q1} \hat{\beta}_{iq0} + \alpha_{q2} \hat{\beta}_{iq1} + \gamma_{qq}\hat{s}_{qq}\right\} + \sum_{q'<q}\gamma_{q'q} \hat{s}_{q'q}.$ For each replicate, we saved the point estimates and 95\% confidence intervals to compute the bias, coverage, and average interval length.

\paragraph{Two-Stage Linear Mixed Model and Linear Regression (TSLMM)}

This alternative is a slightly more sophisticated approach than TSLM. In the first stage, we fit a Bayesian bivariate response linear mixed model with the \textbf{{brms}} package \citep{brms2017} 
\[X_{iq}(t) = \beta_{q0} +  b_{iq0} + \beta_{q1} t + b_{iq1} t + \epsilon_{itq},  q = 1,2.\]

We chose to use a Bayesian framework for this model since fitting linear mixed models with multivariate outcomes is more complicated to implement in a frequentist setting. Standard Bayesian software such as the \textbf{{brms}} package allows for easier implementation of multivariate outcome linear mixed models. We place independent $\mathcal{N}(0, 10^2)$ priors on the intercept and slope parameters. We use the preset prior distribution for the random-effects correlation matrix, an LKJ prior with scale parameter 1, as suggested by \cite{brms2017}. For all other prior specifications, we used the default prior settings in the \textbf{{brms}} package.

We approximated the $``B_{i}"$ coefficients for each individual trajectory with the ``overall" coefficient estimates: $\hat{B}_{iq0}$ =  $\hat{\beta}_{q0} + \hat{b}_{iq0}$ and $\hat{B}_{iq1}$ = $\hat{\beta}_{q1} + \hat{b}_{iq1}$,  where $\hat{\beta}_{qp}$ and $\hat{b}_{iqp}, p =0, 1$ are the estimated posterior means of the fixed and random effects respectively.
As in the previous model, we estimated $\Sb_i$ by computing the model residuals (e.g. $X_{itq} - (\hat{B}_{i0q}+ \hat{B}_{i1q}t$)) and then computed the variance across all residuals. We also computed the residual covariance to estimate of $\hat{s}_{12}$.  We then fit the same second-stage model as in the TSLM setup to get the estimated posterior means and corresponding 95\% confidence intervals for $\balpha$ and $\bgamma$. 

\paragraph{Two-Stage Individual Variances (TSIV) Model} Here, we fit the longitudinal outcome model using Equations (\ref{eq::marker_model}) and (\ref{eq::individual_beta_sigma}) only (together with their prior specifications in Equation (\ref{beta_hyperprior}), (\ref{hyperprior_QiRi}) to (\ref{prior_Bis}))  and use the estimates of the posterior means, $\hat{\Bb}_i$ and $\hat{\Sb}_i$ in the model \ref{eq::outcome_regression} (together with prior specifications \ref{prior_outcome_model}). Note that we do not consider this to be a practical alternative to our first two models, since if one goes to the effort of using a non-standard multilevel model for subject-specific variance-covariance matrices, one might as well go the extra step of bringing them together within a joint model. However, we do this to investigate the effect of not propagating the statistical uncertainty  across the two components of the model.

\subsubsection{Simulation I: Results}
Table \ref{tab:simulation_comparisons_1000_ids} presents the results of Simulation I. For the two-stage linear regression model, we can see that the point estimates of the outcome model parameters are attenuated towards the null. This result makes sense given what we know about bias resulting from measurement error  (Section \ref{sec::measurement_error}). Furthermore, the actual coverage rate is quite poor, especially for the regression coefficients of the variances and covariances ($\bgamma$). 
\begin{table}[!ht]
\centering
\renewcommand{\arraystretch}{0.5}
\begin{tabular}{l l r r r}
\toprule
Truth &Model&{Bias}&{Coverage (\%)}&{Average Interval Length}\\
\midrule
$\alpha_{11}$ = -3&\textbf{{JMIV}}&0.01&98.0&0.29\\
                &TSLM&0.33&2.5&0.30\\
                &TSLMM &-0.01&93.5&0.38\\
                 &TSIV &-0.02&97.5&0.35\\
\addlinespace
$\alpha_{12}$ = -3 &\textbf{{JMIV}}&0.01&95.0&0.27\\
                &TSLM&0.13&67.5&0.34\\
                &TSLMM &0.00&94.0&0.34\\
                 &TSIV &-0.01&93.0&0.29\\
\addlinespace
$\alpha_{21}$ = -3&\textbf{{JMIV}}&0.00&97.0&0.25\\
                &TSLM&0.46&0.0&0.29\\
                &TSLMM &-0.01&92.5&0.40\\
                 &TSIV &-0.01&89.0&0.32\\
\addlinespace
$\alpha_{22}$ = 3&\textbf{{JMIV}}&-0.02&93.5&0.37\\
                &TSLM&-0.17&71.5&0.36\\
                &TSLMM &-0.01&97.0&0.49\\
                 &TSIV &0.01&95.0&0.43\\
\addlinespace
$\gamma_{11}$ = 2 &\textbf{{JMIV}}&0.01&94.0&0.52\\
                &TSLM&-0.38&17.0&0.36\\
                &TSLMM &-0.38&15.0&0.35\\
                 &TSIV &0.03&76.0&0.41\\
\addlinespace
$\gamma_{12}$ = -1&\textbf{{JMIV}}&0.00&95.5&0.86\\
                &TSLM&0.41&35.0&0.65\\
                &TSLMM &-0.41&31.0&0.62\\
                 &TSIV &0.04&88.5&0.86\\
\addlinespace
$\gamma_{22}$ = 2&\textbf{{JMIV}}&0.00&98.0&0.43\\
                &TSLM&0.62&0.0&0.33\\
                &TSLMM &0.62&0.0&0.32\\
                 &TSIV &-0.01&88.5&0.40\\
\bottomrule
\end{tabular}
\caption{Simulation I: bias, coverage, and 95\% credible interval (or confidence interval) length across 200 simulation replicates. We compare our Bayesian joint model (JMIV) to the 1) simple two-stage linear regression (TSLM) 2) the two-stage linear mixed model-linear regression (TSLMM) and 3) the two-stage individual variances (TSIV) model. See Section \ref{sec:alternatives} for details about the alternative methods.}
 \label{tab:simulation_comparisons_1000_ids}
\end{table}

For the TSLMM approach, the coverage and bias of the $\alpha$ parameters have significantly improved compared to the TSLM approach, likely due to the linear mixed model appropriately capturing the dependence between individuals' data points (i.e., appropriately capturing the measurement error in the mean parameters).  However, TSLMM still has difficulty in recovering the coefficients of the variances and covariances, as can be seen by the poor coverage and high bias of these parameters. This makes sense since this framework assumes that individual random effects variability can be drawn from a population level variance-covariance matrix (not capturing measurement error in the variance parameters). This result suggests that if the individual variances and covariances do have an influential role in estimating the outcome, neither TSLMM nor TSLM will be able to recover the true values of these parameters. Interestingly, the TSLMM results also show an attenuation towards the null for the $\bgamma$ parameters, but not for the $\balpha$ parameters (although the bias is negligible). This indicates that the TSLMM alternative is able to better estimate the individual intercepts and slopes, but not the residual variability.

Compared to the TSLM and the TSLMM approaches, the TSIV approach has noticeably better coverage and lower bias of the $\bgamma$ parameters. However, compared to our proposed JMIV, TSIV is still uniformly `worse` across the three metrics. The bias of the three $\bgamma$ parameters is higher when compared to the bias produced by JMIV. Also, none of the $\bgamma$ parameters have higher than 90\% coverage across the 200 replicates and the average length of the 95\% credible intervals is higher than the 95\% credible intervals from the JMIV approach. Across all of the simulation replicates, JMIV achieved greater than 90\% coverage of the true parameters. JMIV also achieved low bias across the simulation replicates. We do note that the average 95\% CrI interval lengths are larger for the $\Rb_i$ parameters than for the $\Db_i$ parameters (see Supplementary Material S5 \citep{chen_supp_2022}). This is likely due to the higher uncertainty in estimating these correlation parameters, which has been captured appropriately. This higher uncertainty is also likely the same mechanism behind the larger average 95\% CrI interval length of the $\gamma_{12}$ parameter (corresponding to the covariance of the two trajectories). Overall, these results demonstrate that our model is able to successfully recover the data generating parameters while maintaining good coverage and low bias.

\subsection{Simulation 2: Three Biomarkers}
\label{sec:ThreeBiomarkersSim}
Here we again simulate $n_i=6$ to $15$ timepoints each for $N=1,000$ individuals. The simulated longitudinal data is generated by 
$X_{itq}  = b_{iq1} + b_{iq2} t + \epsilon_{iq}, q=1,2,$ where 
$\bB_{i1} \sim \mathcal{N}_2\left(\bbeta_{1}, \Sigma_{1}\right),
\bB_{i2} \sim \mathcal{N}_2\left(\bbeta_{2}, \Sigma_{2}\right), 
\bB_{i3} \sim \mathcal{N}_2\left(\bbeta_{3}, \Sigma_{3}\right)$, and $\bbeta_{1}=(0,2)^\transp$, $\bbeta_{2}=(2,1)^\transp$, $\bbeta_{3}=(1,1)^\transp$,
\begin{align*}
&\Sigma_1 =\begin{pmatrix}
1 & -0.05 \\
-0.05 & 1 
\end{pmatrix},
\Sigma_2 = \begin{pmatrix}
1 & -0.1 \\
-0.1 & 0.5 
\end{pmatrix},
 \Sigma_3 =
\begin{pmatrix}
1 & -0.25 \\
-0.25 & 1 
\end{pmatrix}, \\
& \log(d_{i1}) \sim \mathcal{N}(0, 0.75)/2,  
\log(d_{i2}) \sim \mathcal{N}(0.5, 0.5)/2, \log(d_{i3}) \sim \mathcal{N}(0,1)/2.
\end{align*}

We first generate the following values $(c_{12} +1)/2  \sim {\sf Beta}(1,5)$,
$(c_{13} +1)/2  \sim {\sf Beta}(1,5)$,
$(c_{23} +1)/2  \sim {\sf Beta}(2,2)$ and use the approach for $Q=3$ markers described in Section \ref{sec::3biomarker} to generate the individual correlation matrices, $\Rb_i$.

\begin{table}[!ht]
\centering
\renewcommand{\arraystretch}{0.5}
\begin{tabular}{l l r r r}
\toprule
Truth&Model&{Bias}&{Coverage (\%)}&{Average Interval Length}\\
\midrule
$\alpha_{11}$ = -3&\textbf{{JMIV}}&0.01&96.5&0.49\\
                &TSLM&0.61&2.0&0.56\\
                &TSLMM &0.00&93.5&0.69\\
                &TSIV &-0.01&96.0&0.64\\
\addlinespace
$\alpha_{12}$ = -3 &\textbf{{JMIV}}&-0.01&93.0&0.45\\
                &TSLM&0.18&77.5&0.62\\
                &TSLMM &-0.01&94.0&0.62\\
                &TSIV &0.00&94.5&0.55\\
\addlinespace
$\alpha_{13}$ = 3&\textbf{{JMIV}}&-0.01&96.5&0.50\\
                &TSLM&-0.98&0.0&0.55\\
                &TSLMM &0.03&93.0&0.74\\
                &TSIV &-0.01&92.0&0.59\\
\addlinespace
$\alpha_{21}$ = -3&\textbf{{JMIV}}&-0.01&95.0&0.43\\
                &TSLM&0.28&47.0&0.53\\
                &TSLMM &0.003&95.0&0.73\\
                &TSIV &-0.01&96.0&0.81\\
\addlinespace
$\alpha_{22}$ = 3&\textbf{{JMIV}}&-0.01&92.5&0.63\\
                &TSLM&-0.17&71.5&0.36\\
                &TSLMM &0.02&93.5&0.90\\
                &TSIV &0.01&93.5&0.65\\
\addlinespace
$\alpha_{23}$ = 3&\textbf{{JMIV}}&-0.01&97.5&0.47\\
                &TSLM&-0.44&20.5&0.63\\
                &TSLMM &-0.02&91.5&0.65\\
                &TSIV &0.01&96.0&0.59\\
\bottomrule
\end{tabular}
\caption{Simulation II:  bias, coverage, and 95\% credible interval (or confidence interval) length across 200 simulation replicates for the $\balpha$ parameters. We compare our Bayesian joint model (JMIV) to the 1) simple two stage linear regression (TSLM) 2) the two stage linear mixed model-linear regression (TSLMM) and 3) the Bayesian two stage model (TSIV).}
 \label{tab:sim2_model_comparisons_alpha}
\end{table}

\begin{table}[!ht]
\centering
\renewcommand{\arraystretch}{0.5}
\begin{tabular}{l l r r r}
\toprule
Truth Parameter&Model&{Bias}&{Coverage (\%)}&{Average Interval Length}\\
\midrule
$\gamma_{11}$ = 2 &\textbf{{JMIV}}&-0.01&93.5&1.08\\
                &TSLM&-0.43&47.5&0.82\\
                &TSLMM &-0.44&43.0&0.80\\
                &TSIV &-0.02&83.0&0.97\\
\addlinespace
$\gamma_{12}$ = -1&\textbf{{JMIV}}&-0.04&92.0&1.66\\
                &TSLM&1.22&8.5&1.32\\
                &TSLMM &1.22&8.0&1.28\\
                &TSIV &-0.08&91.5&1.77\\
\addlinespace
$\gamma_{22}$ = 2&\textbf{{JMIV}}&0.03&96.5&0.79\\
                &TSLM&-1.23&0.0&0.61\\
                &TSLMM &-1.24&0.0&0.59\\
                &TSIV &-0.03&87.5&0.76\\
\addlinespace
$\gamma_{13}$ = -2&\textbf{{JMIV}}&0.03&94.5&1.91\\
                &TSLM&1.63&2.5&1.33\\
                &TSLMM &1.62&2.5&1.29\\
                &TSIV &-0.06&82.0&1.78\\
\addlinespace
$\gamma_{23}$ = 2&\textbf{{JMIV}}&0.04&94.5&1.50\\
                &TSLM&-0.90&17.5&1.07\\
                &TSLMM &-0.92&14.5&1.03\\
                &TSIV &-0.01&90.5&1.52\\
\addlinespace
$\gamma_{33}$ = 1&\textbf{{JMIV}}&0.01&94.0&0.54\\
                &TSLM&0.01&68.5&0.38\\
                &TSLMM &-0.002&69.0&0.37\\
                &TSIV &-0.02&64.4&0.42\\
\bottomrule
\end{tabular}
\caption{Simulation II: bias, coverage, and 95\% credible interval (or confidence interval) length across 200 simulation replicates for the $\bgamma$ parameters. We compare our Bayesian joint model (JMIV) to the 1) simple two stage linear regression (TSLM) 2) the two stage linear mixed model-linear regression (TSLMM) and 3) the Bayesian two stage model (TSIV).}
 \label{tab:sim2_model_comparisons_gamma}
\end{table}

\par For the outcome submodel, we set the true values of the regression coefficients as: $\balpha = (-3,-3, 3,-3,3, 3), \bgamma = (2,-1,2, -2, 2, 1)$, where these values were again chosen so that the distribution of the outcome $y_i$ would be similar to the distribution of the SWAN body mass outcomes. Lastly, we set $\sigma^2 = 1$ (the variance parameter for the outcome). 

We present the results of this simulation study in Tables \ref{tab:sim2_model_comparisons_alpha} and \ref{tab:sim2_model_comparisons_gamma}. We note that the proposed JMIV achieves above 90\% coverage for both the mean ($\balpha$) and variance-covariance ($\bgamma$) parameters, which the other models fail to do. With respect to the $\balpha$ parameters, TSLMM and TSIV both perform well in terms of both coverage and bias. However, substantial differences in performance are present in the $\bgamma$ parameters. We see that that our model, JMIV, consistently has lower bias except in the case of $\gamma_{23}$ where TSIV achieves lower bias and $\gamma_{33}$ where TSLMM achieves lower bias. However, in both cases, JMIV outperforms the other models in terms of higher coverage (substantially higher coverage in the case of $\gamma_{33}$), indicating that JMIV is still a better model choice.

\section{Hormone Trajectories and Changes in Body Mass Across the Menopausal Transition}
\label{sec::swan_application}
\subsection{Study of Women's Health Across the Nation}
The Study of Women's Health Across the Nation (SWAN) is a longitudinal cohort study that followed women over the menopausal transition \citep{sowers_chapter_2000}. Seven clinical sites across the United States participated in this study. To be eligible for the cohort, women had to be between 42-52 years old, had to have had at least one menstrual period and not used reproductive hormones (e.g. hormonal contraceptives or other exogenous hormones) in the past three months, had to reside in the geographic area of the clinical site, and had to self-identify as White, Black, Chinese, Japanese or Hispanic. Serum E2 and FSH biomarker measurements were collected at baseline and during 13 approximately annual follow-up visits, along with other health measurements. Body composition was measured via dual-energy X-ray absorptiometry (DXA) at five of the seven clinical site visits. Women also completed questionnaires regarding lifestyle and socio-demographic characteristics. The comprehensive and longitudinal aspect of this dataset makes it an ideal application for understanding how individual hormone trajectories affect the rate of change of fat mass and lean mass across the menopausal transition.

\par We initially started with women who were enrolled at one of the five sites with body composition measures and who were not on hormone therapy during the clinical visits where E2 and FSH measurements were taken. After restricting our analysis to women who had an observed FMP, we had 887 individuals in the dataset. Although there are five racial/ethnic groups in the SWAN study, the site with Hispanic women did not have body composition data; hence, Hispanic women are not included in this analysis.  After computing the body mass composition windows (see below), an additional 47 women were deleted from the analysis. The final analytical dataset was completed on 841 individuals, with a total number of 9,902 hormone measurements. 

\begin{figure}
    \centering
    \includegraphics[scale=0.2]{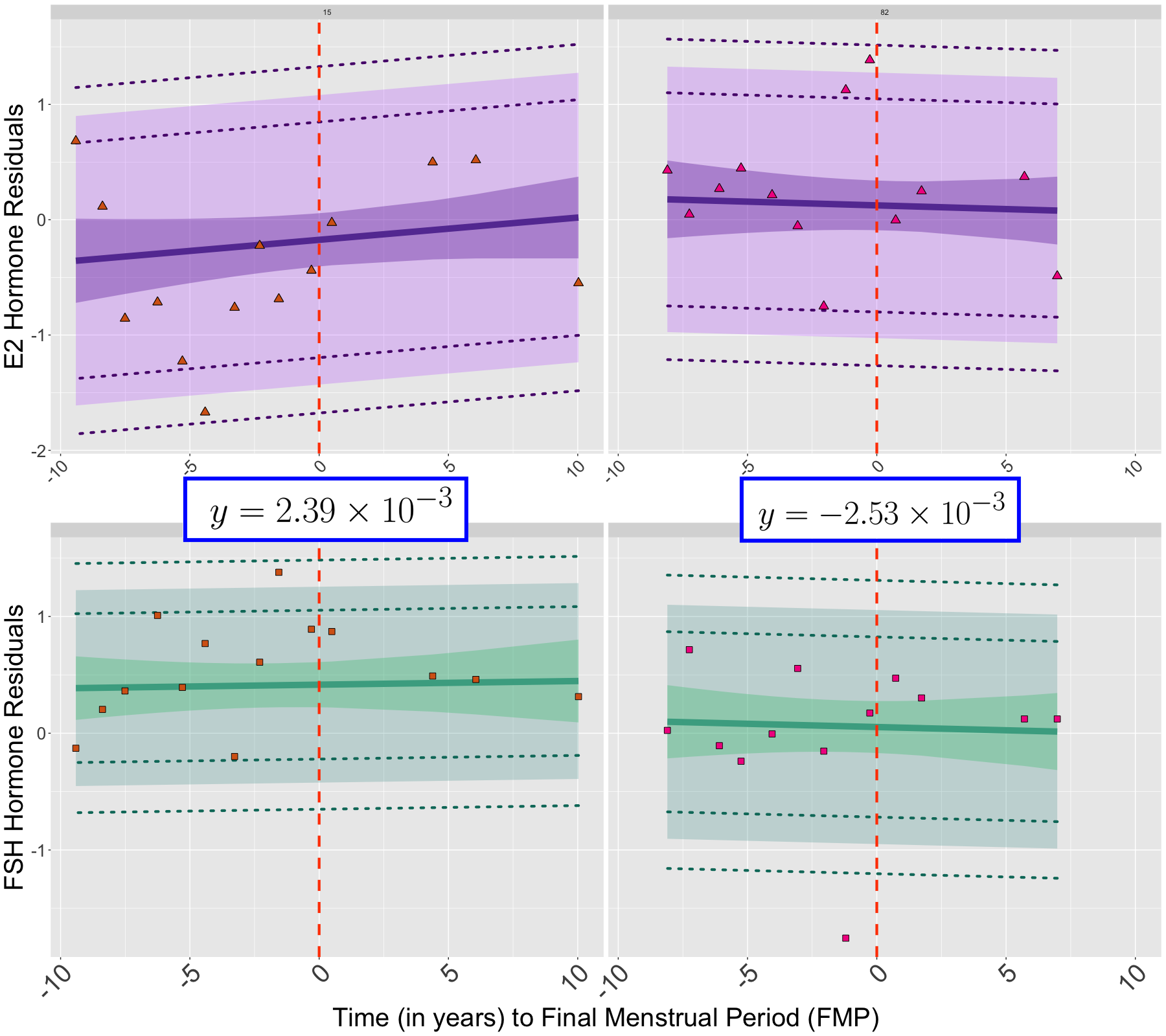}
    \caption{Plots of estimated hormone trajectories for two individuals from the fat mass rate of change model. The solid lines are the estimated individual mean trajectories, based on the posterior means of \boldsymbol{$B_i$},  i.e. $\hat{b}_{i0} + \hat{b}_{i1} t$. The darker inner intervals around the  solid lines are $+/- 1.64 \times var(\hat{b}_{i0} + \hat{b}_{i1} t)$ and the lighter band is $+/- 1.64 \times \hat{\sigma}_{iq}$, where $\hat{\sigma}_{iq}$ is the square root of the estimated posterior mean of the individual level variance of hormone $q$. The dotted lines represent $+/- 1.64 \times \sigma_{iq5}$ and $+/- 1.64 \times \sigma_{iq95}$ where $\sigma_{iq5}, \sigma_{iq95}$ are the values of the 5th and 95th percentiles of the posterior samples of the individual variances for each hormone $q$. The triangles and squares are the observed E2 and FSH residuals, respectively. The observed individual fat mass rates of change are shown in bordered boxes.}
    \label{fig:plot_2_individuals}
\end{figure}

For the individual trajectory model, we use the log values of FSH (mIU/mL) and E2 (pg/mL) measured at each visit. The log scale minimizes the skewness of the data and allows us to better determine the overall population trend. We removed this population trend by fitting a lowess curve to each (log) hormone. This was done by using time to FMP at each visit as the numeric predictor for the corresponding hormone measurement and using weighted least squares to obtain the predicted fit at each timepoint. We then subtracted the individual measurements from the lowess estimates. By removing the common population trends in the data,  our model can better approximate the individual trajectories and individual level variances using a simpler (lower dimensional) subject-level trajectory model. Figure \ref{fig:plot_2_individuals} summarizes the subject-level longitudinal data model fit results for two randomly-selected women. These individual plots in Figure \ref{fig:plot_2_individuals} indicate that our model generates reasonable estimates of the individual hormone values.

\begin{figure}
    \centering
    \includegraphics[scale=0.2]{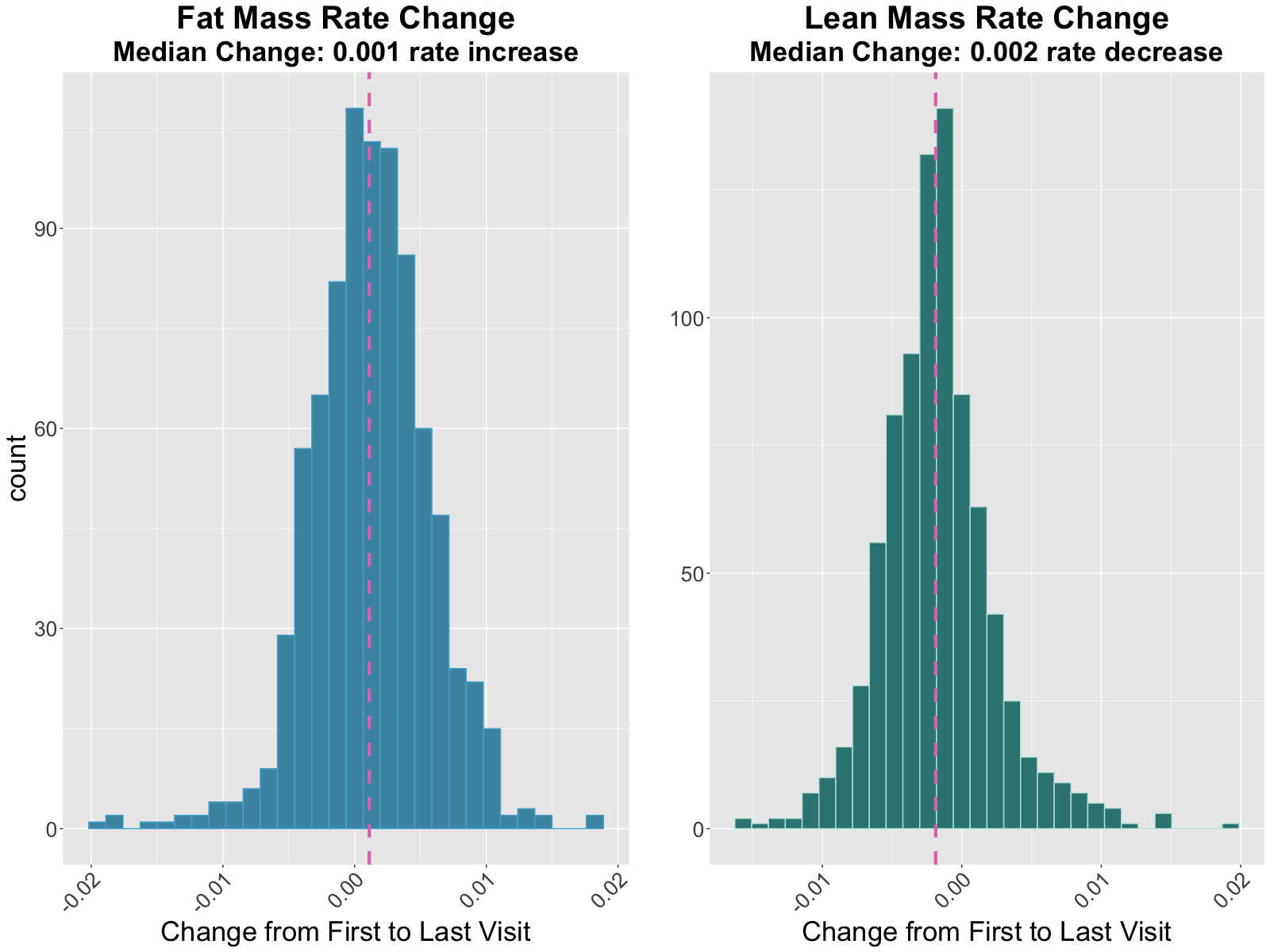}
    \caption{Histograms of the observed rate of change in fat mass composition (left) and the observed rate of change in lean mass composition (right).}
    \label{fig:bm_histograms_2x2}
\end{figure}

For the outcomes of interest, we selected fat mass rate of change and lean mass rate of change over a selected window. We define this window to be from the visit closest to 5 years before the FMP to the visit closest to 5 years after the FMP, with the requirement that the closest visit be at least 3 or more years before/after the FMP. By doing this, we aimed to capture the most accurate trend in body composition change that was not fully dependent on a measurement right before menopause. 
The fat mass (or lean mass) rate of change is the difference between the `first visit' (pre FMP) and the `last visit' (post FMP) divided by the amount of time (in years) within each individual window. We normalized the body mass measurements by using the proportion of fat mass (or lean mass) to body weight (grams) rather than using the unadjusted body mass measurements (both measured in grams).  Figure \ref{fig:bm_histograms_2x2} displays the histograms of these outcomes, after performing the normalization and rate adjustments. 

In the outcome models, we used the correlation between E2 and FSH rather than the covariance as a variable of interest, since the correlation measure has a more straightforward interpretation and is normalized to the E2, FSH variances. We also explored interaction terms in the outcome regression, in order to better understand the relationships between the two hormones, and how these may affect the body mass outcomes. We interacted the individual E2 and FSH intercepts, individual E2 and FSH slopes, and individual E2 and FSH variances parameters separately for both the fat mass and lean mass outcomes (for a total of six different models). For both the fat mass and lean mass models, significant interactions between the E2 and FSH variances were observed. 

Finally, by including additional covariates  $\bW_i$ in the outcome regression model, we adjusted for: fat mass (lean mass) body weight proportion at the `first' visit, race/ethnicity (White, Black, Chinese and Japanese) and sports activity category. We included race/ethnicity in the models given previous research using SWAN data that found differences in body mass composition changes among ethnic groups during the menopausal transition \citep{greendale_changes_2019,greendale_changes_2021}. The physical activity category is a measure of the individual level physical activity trajectories for each subject in the SWAN study, grouped into categories reflecting: (1) lowest, (2) increasing, (3) decreasing, (4) middle, and (5) highest physical activity during follow up. For a more detailed description, please refer to \cite{pettee_gabriel_physical_2017}. Table \ref{tab:swan_descriptive_stats} displays the descriptive statistics for the individuals in our analysis, including demographic and physical activity information. The following sections contain the results from applying our joint model to the SWAN dataset.

\begin{table}
\centering
\renewcommand{\arraystretch}{0.3}
\begin{tabular}{l r r r}
\toprule
\textbf{{Variable}}&\textbf{Statistic}&\textbf{Value}&\textbf{{n}}\\
\midrule
\textit{Longitudinal Predictors} & Mean/SD &&\\
E2 Residuals & &  -0.04 (0.81) & 9,902 \\
FSH Residuals & & 0.02 (0.61) & 9,902 \\
\addlinespace 
\textit{Body Mass Outcome} & Mean/SD &&\\
Fat Mass Rate of Change  & & 0.001 (0.004) & 841 \\
Lean Mass Rate of Change & &  -0.001 (0.004) & 841 \\
\addlinespace 
\textit{Baseline Body Mass} & Mean/SD &&\\
Fat Mass Prop. at Visit 1 & & 0.36 (0.07) & 841 \\
Lean Mass Prop. at Visit 1 & & 0.55 (0.06) & 841 \\
\addlinespace
\textit{Race/Ethnicity} & Percent &&\\
White (Reference) & &  47.2\% & 397  \\
Black&  & 24.9\% & 209  \\
Japanese&  & 15.3\% & 129 \\
 Chinese& &  12.6\% & 106  \\
\addlinespace
{\it Physical Activity}& Percent && \\
Lowest Activity (Reference) &  &  23.6\% & 199  \\
Increasing Activity & & 12.7\% & 107  \\
Decreasing Activity & & 22.7\% &191 \\
Middle Activity & & 25.6\% & 215  \\
Highest Activity & & 15.3\% & 129  \\
\bottomrule
\end{tabular}
\caption{Descriptive statistics of the SWAN dataset based on 840 individuals.}
\label{tab:swan_descriptive_stats}
\end{table}

\subsection{Fat Mass Rate of Change}

Table \ref{tab:fat_mass_var_interaction} displays the results of the fat mass model with variance interactions. We found that the E2, FSH intercepts and the E2, FSH slopes were all signfiicantly associated with fat mass rate of change. For a one-unit increase in starting E2 (from the population average), this was associated with an average $5.02 \times 10^{-3}$ increase in fat mass per year. 

Since we standardized the hormones to be relative to the population average, the effects of the E2 and FSH variances can be interpreted as the effect of one hormone's variability when the other hormone variability is at the population average. So, when FSH variability was the same as the population variance, then a one unit increase in E2 variability was associated with a $7.03 \times 10^{-3}$ increase in fat mass per year. Similarly, when E2 variability was the same as the population variance, then a one unit increase in FSH variability was associated with a $16.38 \times 10^{-3}$ inrease in fat mass per year. 

\begin{table}
\centering
\renewcommand{\arraystretch}{0.2}
   \begin{tabular}{l r r r r r}
\toprule
\multicolumn{1}{l}{Coefficient}&\multicolumn{1}{c}{Variable}&\multicolumn{1}{c}{Post. Mean}&\multicolumn{1}{c}{2.5\% CrI}&\multicolumn{1}{c}{97.5\% CrI}\tabularnewline
\midrule
$\alpha_{11}$ & \textbf{E2 Intercept} & \textbf{5.02}& \textbf{2.56}& \textbf{7.77}& \tabularnewline
$\alpha_{21}$ & \textbf{FSH Intercept} & \textbf{0.94}& \textbf{-0.20} & \textbf{2.09} &\tabularnewline
$\alpha_{12}$& \textbf{E2 Slope} & \textbf{33.70}& \textbf{4.28}& \textbf{68.54}&\tabularnewline
$\alpha_{22}$ & \textbf{FSH Slope} & \textbf{-23.82} & \textbf{-42.83} & \textbf{-4.82}&\tabularnewline
$\gamma_{11}$ & \textbf{E2 Var.}  &\textbf{7.03} & \textbf{2.63} & \textbf{10.92} &\tabularnewline
$\gamma_{12}$ & E2, FSH Cor.   &2.44 & -1.51 & 6.11 &\tabularnewline
$\gamma_{22}$ &  \textbf{FSH Var.}   &\textbf{16.38} & \textbf{5.43}& \textbf{26.42} &\tabularnewline
$\gamma_{3}$ &  \textbf{E2 Var. x FSH Var.}  &\textbf{-33.78} & \textbf{-51.62} & \textbf{-14.47} &\tabularnewline
$\gamma^{W}_{1}$ & \textbf{Fat Mass Prop. (First Visit)} & \textbf{-2.05} & \textbf{-2.38} & \textbf{-1.74} &\tabularnewline
$\gamma^{W}_{2}$& \textbf{Black}  & \textbf{-0.73}& \textbf{-1.42} & \textbf{0.00} &\tabularnewline
$\gamma^{W}_{3}$ &Chinese & 0.11 & -0.80 & 0.98 &\tabularnewline
$\gamma^{W}_{4}$ & \textbf{Japanese} & \textbf{-3.23} & \textbf{-4.07}& \textbf{-2.41} &\tabularnewline
$\gamma^{W}_{5}$ & Increasing Activity (Cat. 2) & 0.30& -0.64 & 1.26  &\tabularnewline
$\gamma^{W}_{6}$ & Decreasing Activity (Cat. 3)& 0.66  &-0.12 & 1.45 &\tabularnewline
$\gamma^{W}_{7}$ & Middle Activity (Cat. 4) & 0.08 & -0.69 & 0.84 &\tabularnewline
$\gamma^{W}_{8}$ &Highest Activity (Cat. 5) & -0.76 &   -1.66 & 0.18  & \tabularnewline

\bottomrule
\end{tabular}
    \caption{Estimated posterior means and 95\% credible intervals for the fat mass rate of change model with variance interactions. All estimated posterior means and 95\% CrI values have been multiplied by $10^{3}$.}
    \label{tab:fat_mass_var_interaction}
\end{table}

\begin{figure*}[t!]
    \centering
    \begin{subfigure}
        \centering
        \includegraphics[scale=0.2]{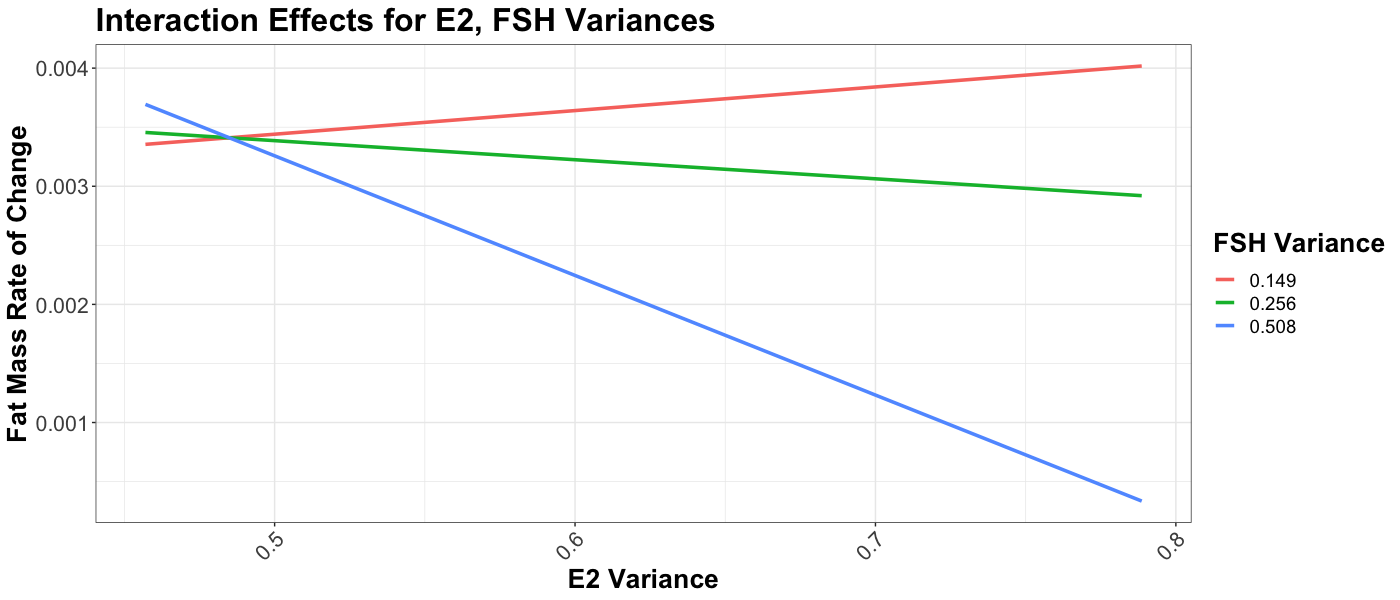}
    \end{subfigure}
    \begin{subfigure}
        \centering
        \includegraphics[scale=0.2]{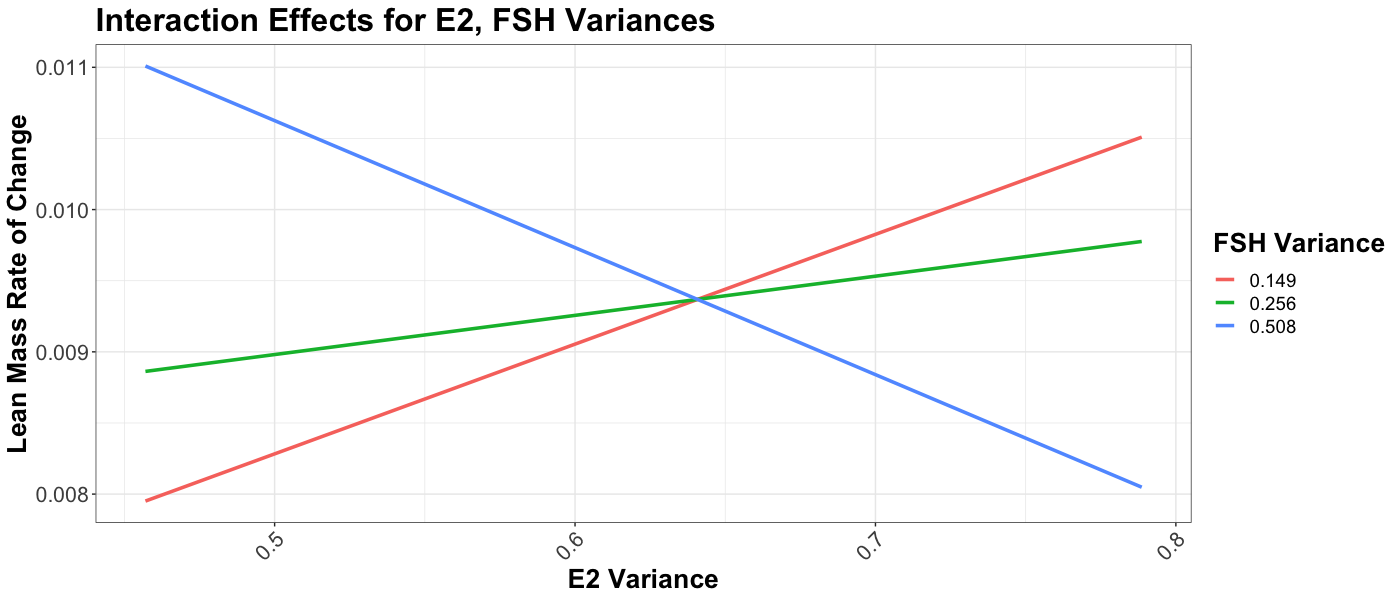}
    \end{subfigure}
    \caption{Visualizations of the effects of the E2, FSH variance terms for the fat mass model (top) and lean mass model(bottoms).}
    \label{fig::var_interactions}
\end{figure*}

The top plot in Figure \ref{fig::var_interactions} shows a visualization of how the interaction between E2, FSH variances is related to fat mass change. We can see that when E2 variability was high and FSH variability was low, there was an average higher fat mass rate of change. However, when both E2 and FSH variability were high, their individual effects on fat mass change were moderated by the interaction term, resulting in lower fat mass rates of change. This finding suggested that women who had highly variable E2 and FSH also had, on average, slower fat mass increases than women who had only one highly variable hormone. E2 and FSH have a tightly coupled relationship since the two hormones balance each other; thus, if one hormone is highly variable when the other one is not, this could indicate higher deterioration or malfunction in the reproductive system's functioning beyond the expected decline during the menopausal transition. If this is the case, then it would make sense that a greater decline in reproductive functioning may be associated with increased fat mass gain, since E2 is known to regulate adipose tissue. 

 In this fully adjusted model, fat mass as a proportion of overall body mass at the first visit was also negatively associated with the outcome (an average $-2.05 \times 10^{-3}$ fat mass decline per year). Black women and Japanese women,  compared to white women, had slower fat mass gain ($-0.73 \times 10^{-3}$ per year and $-3.23 \times 10^{-3}$ per year, respectively).

\subsection{Lean Mass Rate of Change}
\label{sec::mixture_model_swan}

Initially the Gaussian outcome assumption did not appear to be a good fit for the observed outcome. In particular, the residuals suggested overdispersed variances with a common mean, so we allowed the outcome to be modeled as a mixture of two Gaussian distributions with equal means but different variances: 
\begin{align*}
      &  Y_i|z_i, \eta(\Bb_{i}, \Sb_{i}, \bW_i), \sigma^2_{1}, \sigma^2_{2} \sim \mathcal{N}(\eta_i,\sigma^2_{z_i}),\\
 &   z_i | \pi \sim{\sf Bernoulli}(\pi),\\
  &  \pi \sim {\sf Beta}(1/2, 1/2), \sigma_{1} \sim {\sf half-Cauchy}(0, 2.5), \sigma_{2} \sim {\sf half-Cauchy}(0, 5)
 \end{align*} where  $z_i$ is an unobserved indicator variable indicating membership in the first mixture component; $z_i=1$ for the first component. Because the mean is equal across the mixture components, the interpretation of the regression parameters will be the same as for the fat mass models despite the additional variance parameter.
 
 Table \ref{tab:lean_mass_var_interaction} displays the estimated coefficients for the lean mass model with variance interactions. A one-unit higher E2 intercept (above the population mean) was associated with an average $-2.54 \times 10^{3}$ decline in lean mass per year. A higher individual FSH intercept was also negatively associated with lean mass; a one-unit higher starting FSH was associated with an average $-1.49 \times 10^{3}$ decline in lean mass per year. 
 
 We found that when the individual E2 variance was at the population mean level, a one unit increase in FSH variance was associated with a $29.70 \times 10^{3}$ increase in lean mass per year. The interaction term of E2, FSH variances was also negative, which indicates that the positive relationship between higher E2 variance and lean mass rate of change was weaker when FSH variability is also high (and vice versa for the relationship between FSH variance and lean mass change). This also corresponds with the E2, FSH correlation term, which is negatively associated with the outcome (although in this model, the 95\% CrI contains 0 so the correlation is not significantly associated with the outcome).

\begin{table}
\centering
\renewcommand{\arraystretch}{0.2}
   \begin{tabular}{l r r r r r}
\toprule
\multicolumn{1}{l}{Coefficient}&\multicolumn{1}{c}{Variable}&\multicolumn{1}{c}{Post. Mean}&\multicolumn{1}{c}{2.5\% CrI}&\multicolumn{1}{c}{97.5\% CrI}\tabularnewline
\midrule
$\alpha_{11}$ & \textbf{E2 Intercept} & \textbf{-2.54}& \textbf{-4.74}& \textbf{-0.40}& \tabularnewline
$\alpha_{21}$ & \textbf{FSH Intercept}& \textbf{-1.49}& \textbf{-2.47} & \textbf{-0.59} &\tabularnewline
$\alpha_{12}$& E2 Slope & -18.46& -46.02& 6.65&\tabularnewline
$\alpha_{22}$ & FSH Slope & 16.27 & -1.02 & 33.40&\tabularnewline
$\gamma_{11}$ & \textbf{E2 Var.}  &\textbf{14.62} & \textbf{9.56} & \textbf{20.27} &\tabularnewline
$\gamma_{12}$ & E2, FSH Cor.   &-3.53 & -7.67 & 0.14 &\tabularnewline
$\gamma_{22}$ &  \textbf{FSH Var.}   &\textbf{29.70} & \textbf{18.17}& \textbf{41.88} &\tabularnewline
$\gamma_{3}$ &  \textbf{E2 Var. x FSH Var.}  &\textbf{-46.35} & \textbf{-67.53} & \textbf{-26.88} &\tabularnewline
$\gamma^{W}_{1}$ & \textbf{Lean Mass Prop. (First Visit)} & \textbf{-24.92} & \textbf{-29.69} & \textbf{-19.89} &\tabularnewline
$\gamma^{W}_{2}$& \textbf{Black}  & \textbf{0.86} & \textbf{0.26} &\textbf{1.48} &\tabularnewline
$\gamma^{W}_{3}$ &\textbf{Chinese}  & \textbf{1.59} & \textbf{0.81} & \textbf{2.36} &\tabularnewline
$\gamma^{W}_{4}$ & Japanese & 0.51 & -0.19& 1.22 &\tabularnewline
$\gamma^{W}_{5}$ & Increasing Activity (Cat. 2) & -0.04& -0.83 & 0.77  &\tabularnewline
$\gamma^{W}_{6}$ & Decreasing Activity (Cat. 3)& -0.09  &-0.75  &0.62 &\tabularnewline
$\gamma^{W}_{7}$ & Middle Activity (Cat. 4) & 0.16 &-0.51 & 0.85 &\tabularnewline
$\gamma^{W}_{8}$ &\textbf{ Highest Activity (Cat. 5)} & \textbf{1.02} & \textbf{0.23} & \textbf{1.81} & \tabularnewline

\bottomrule
\end{tabular}
    \caption{Estimated posterior means and 95\% credible intervals for the lean mass rate of change model with variance interactions. All estimated posterior means and 95\% CrI values have been multiplied by $10^{3}$.}
    \label{tab:lean_mass_var_interaction}
\end{table}

The bottom plot of Figure \ref{fig::var_interactions} displays how the variance interaction affects lean mass change for different values of E2 variance and FSH variance. When E2 variability was high and FSH variability was low, there was an average higher lean mass rate of change. This was also true when FSH variability was high and E2 variability was low. However, when both E2 and FSH variabilities were high, their effects on lean mass change was  moderated by the interaction term, resulting in average lower lean mass changes. Women who had highly variable E2 and FSH also had, on average, slower lean mass increases than women who had one, but not both, highly variable hormone. This result mirrors the result from the fat mass rate of change model, where we also saw that the variance interaction term moderated the effects from the E2 and FSH variance terms. 

Unsurprisingly, lean mass proportion at first visit was negatively associated with lean mass change ($-24.92 \times 10^{3}$ decline per year). Black and Chinese women also had faster lean mass gains compared to white women on average ($0.86 \times 10^{3}$ increase per year and $1.59 \times 10^{3}$ increase per year, respectively). Finally, women in the highest activity group also had higher lean mass increases than women in the lowest activity group ($1.02 \times 10^{3}$ increase per year).

\section{Discussion}
\label{sec::discussion}
We have presented a joint modeling approach for estimating individual-level mean and variance-covariance matrices based on longitudinal marker trajectories, which are then linked to a cross-sectional outcome.  Simulations show that our model outperforms alternative approaches to this research problem. Analysis of hormone trajectories data reveal that the variability of E2 does have a statistically significant association with the rate of change of body mass composition across the menopausal transition.

Our work is important for both methodological development of joint models and for women's health research. Our model estimates both mean longitudinal trends and the residual variability of these individual trajectories, and propagates the estimation uncertainty into the second submodel. This joint modelling is important for obtaining accurate estimates (in terms of low bias, higher coverage and lower variability) of how the individual-level parameters are linked to the outcome. Simulation results demonstrate that our model outperforms common two-stage approaches. Substantively, understanding how the individual-level variability of two hormones (E2 and FSH) is associated with changes in body mass composition had not been well explored. Our analysis of the SWAN data revealed that higher variability in E2 and FSH are linked to both faster rates of fat mass and lean mass change. Our analysis of the SWAN data revealed that higher variability in E2 and FSH are linked to both faster rates of fat mass and lean mass change. Additionally, women with highly variable E2 and FSH tended to have slower fat mass increases and slower lean mass increases than women who had only one highly variable hormone. These findings can serve as a basis for further investigation of how hormone variability affects such health outcomes. 

As mentioned above, joint estimation of longitudinal variables and scalar outcomes can be useful for understanding of scientific questions in many areas. With longitudinal biomarker data becoming more readily available (e.g. wearable devices), we need statistical methods for analyzing these types of data. Our proposed method addresses the gap in methods by 1) providing a framework for jointly modeling longitudinal and cross-sectional data and 2) explicitly modeling individual-level variability in the longitudinal trajectories, which can improve understanding of the relationship between longitudinal predictors and health outcomes. 

\subsection{Remark} In our simulation studies and scientific application, we made the simplifying assumption to exclude covariates $\bW_i$ in modeling the longitudinal markers (Equation \ref{eq::marker_model}). In mathematical terms, this means that the likelihood functions of $\bX_{it}, \bB_i, \bS_i, \bW_i$ are: 

\begin{align*}
    f(\bX_{it} \mid \bB_i, \bS_i, \bW_i) =  f(\bX_{it} \mid \bB_i, \bS_i) \\
     f(\bB_i, \bS_i \mid \bW_i) =  f(\bB_i, \bS_i) \\
\end{align*}
 For the scientific application, our main focus was to evaluate the overall marginal effects of the biomarker hormone means and variances on the body mass outcomes of interest. In particular, the effects of the variance and correlation parameters were of key interest, since the associations between individual E2 and FSH variabilities (and co-variability) and body mass changes had not been previously explored. The estimated coefficients for the $\bB_i, \bS_i$ described in Section \ref{sec::swan_application} should be interpreted as marginal effects, rather than conditional on the other adjusted covariates. For this particular scientific application, we believed that this assumption resulted in a more straightforward interpretation of the mean, variance, and correlation parameters. The decision to include or exclude $\bW_i$ in the longitudinal submodel should be made with the specific research application in mind and whether or not the simplifying assumption makes sense for the particular context.

\subsection{Future Work} 

One extension of this work could be to model the individual variances as being functions of time, i.e. $\Sb_{it}$.  E2 and FSH are known to be highly variable as women approach their final menstrual cycle, so estimating $\Sb_{it}$ may better capture such changes in the biomarker variances. To obtain these estimates, we would likely need a larger dataset (with both more individuals and timepoints) than is currently available with the SWAN study. Another methodological extension could be to extend this model to account for missingness in both the trajectory data and the outcome data. For this analysis, we removed the missing values in the hormone data and only analyzed individuals with observed body mass outcomes. Although less than 5\% of the values in our dataset were missing, analyzing complete case data only could still result in biased inference. In the SWAN dataset, individuals can be subject to intermittent missingness as well as dropout; these types of missing data patterns could be addressed in future work. Finally, an alternative approach could extend the work of \citep{jiang_joint_2015} in using latent class models to capture the impact of the means and variances of multiple longitudinal trajectories on health outcomes.
Finally, we note that increasing the number of longitudinal trajectories may result in a form of $\eta(\cdot)$ in the OR model that is complicated to estimate, since the number of variance-covariance parameters increases quadratically with the number of trajectories. Some type of dimension reduction procedure may be useful in these settings, although retaining interpretability may be challenging.

\bibliographystyle{unsrtnat}
\bibliography{references}  

\begin{thebibliography}{41}
\providecommand{\natexlab}[1]{#1}
\providecommand{\url}[1]{\texttt{#1}}
\expandafter\ifx\csname urlstyle\endcsname\relax
  \providecommand{\doi}[1]{doi: #1}\else
  \providecommand{\doi}{doi: \begingroup \urlstyle{rm}\Url}\fi

\bibitem[Randolph et~al.(2004)Randolph, Sowers, Bondarenko, Harlow, Luborsky,
  and Little]{randolph_change_2004}
John~F. Randolph, Jr., {MaryFran} Sowers, Irina~V. Bondarenko, Siobán~D.
  Harlow, Judith~L. Luborsky, and Roderick~J. Little.
\newblock Change in estradiol and follicle-stimulating hormone across the early
  menopausal transition: Effects of ethnicity and age.
\newblock \emph{The Journal of Clinical Endocrinology \& Metabolism},
  89:\penalty0 1555--1561, 2004.
\newblock ISSN 0021-972X.
\newblock \doi{10.1210/jc.2003-031183}.
\newblock URL \url{https://doi.org/10.1210/jc.2003-031183}.

\bibitem[Karvonen-Gutierrez and
  Harlow(2017)]{karvonen-gutierrez_menopause_2017}
Carrie Karvonen-Gutierrez and Sioban~D. Harlow.
\newblock Menopause and midlife health changes.
\newblock In Jeffrey~B. Halter, Joseph~G. Ouslander, Stephanie Studenski,
  Kevin~P. High, Sanjay Asthana, Mark~A. Supiano, and Christine Ritchie,
  editors, \emph{Hazzard's Geriatric Medicine and Gerontology}. {McGraw}-Hill
  Education, 7 edition, 2017.
\newblock URL \url{accessmedicine.mhmedical.com/content.aspx?aid=1136589663}.

\bibitem[Colleluori et~al.(2018)Colleluori, Chen, Napoli, Aguirre, Qualls,
  Villareal, and Armamento-Villareal]{colleluori_fat_2018}
Georgia Colleluori, Rui Chen, Nicola Napoli, Lina~E. Aguirre, Clifford Qualls,
  Dennis~T. Villareal, and Reina Armamento-Villareal.
\newblock Fat mass follows a u-shaped distribution based on estradiol levels in
  postmenopausal women.
\newblock \emph{Frontiers in Endocrinology}, 9:\penalty0 315, 2018.
\newblock ISSN 1664-2392.
\newblock \doi{10.3389/fendo.2018.00315}.
\newblock URL \url{https://www.ncbi.nlm.nih.gov/pmc/articles/PMC6036116/}.

\bibitem[Randolph et~al.(2011)Randolph, Zheng, Sowers, Crandall, Crawford,
  Gold, and Vuga]{randolph_change_2011}
John~F. Randolph, Huiyong Zheng, {MaryFran}~R. Sowers, Carolyn Crandall, Sybil
  Crawford, Ellen~B. Gold, and Marike Vuga.
\newblock Change in follicle-stimulating hormone and estradiol across the
  menopausal transition: Effect of age at the final menstrual period.
\newblock \emph{The Journal of Clinical Endocrinology and Metabolism},
  96:\penalty0 746--754, 2011.
\newblock ISSN 0021-972X.
\newblock \doi{10.1210/jc.2010-1746}.
\newblock URL \url{https://www.ncbi.nlm.nih.gov/pmc/articles/PMC3047231/}.

\bibitem[Zaidi et~al.(2018)Zaidi, Lizneva, Kim, Sun, Iqbal, New, Rosen, and
  Yuen]{zaidi_fsh_2018}
Mone Zaidi, Daria Lizneva, Se-Min Kim, Li~Sun, Jameel Iqbal, Maria~I New,
  Clifford~J Rosen, and Tony Yuen.
\newblock {FSH}, bone mass, body fat, and biological aging.
\newblock \emph{Endocrinology}, 159:\penalty0 3503--3514, 2018.
\newblock ISSN 0013-7227.
\newblock \doi{10.1210/en.2018-00601}.
\newblock URL \url{https://www.ncbi.nlm.nih.gov/pmc/articles/PMC6134257/}.

\bibitem[Sponton and Kajimura(2017)]{sponton_burning_2017}
Carlos~Henrique Sponton and Shingo Kajimura.
\newblock Burning fat and building bone by {FSH} blockade.
\newblock \emph{Cell Metabolism}, 26:\penalty0 285--287, 2017.
\newblock ISSN 1932-7420.
\newblock \doi{10.1016/j.cmet.2017.07.018}.

\bibitem[Kohrt and Wierman(2017)]{kohrt_preventing_2017}
Wendy~M. Kohrt and Margaret~E. Wierman.
\newblock Preventing fat gain by blocking follicle-stimulating hormone.
\newblock \emph{The New England Journal of Medicine}, 377:\penalty0 293--295,
  2017.
\newblock ISSN 1533-4406.
\newblock \doi{10.1056/NEJMcibr1704542}.

\bibitem[Gourlay et~al.(2012)Gourlay, Specker, Li, Hammett-Stabler, Renner, and
  Rubin]{gourlay_follicle-stimulating_2012}
Margaret~L. Gourlay, Bonny~L. Specker, Chenxi Li, Catherine~A. Hammett-Stabler,
  Jordan~B. Renner, and Janet~E. Rubin.
\newblock Follicle-stimulating hormone is independently associated with lean
  mass but not {BMD} in younger postmenopausal women.
\newblock \emph{Bone}, 50:\penalty0 311--316, 2012.
\newblock ISSN 8756-3282.
\newblock \doi{10.1016/j.bone.2011.11.001}.
\newblock URL \url{https://www.ncbi.nlm.nih.gov/pmc/articles/PMC3246561/}.

\bibitem[Liu et~al.(2017)Liu, Ji, Yuen, Rendina-Ruedy, {DeMambro}, Dhawan,
  Abu-Amer, Izadmehr, Zhou, Shin, Latif, Thangeswaran, Gupta, Li, Shnayder,
  Robinson, Yu, Zhang, Yang, Lu, Zhou, Zhu, Oberlin, Davies, Reagan, Brown,
  Kumar, Epstein, Iqbal, Avadhani, New, Molina, van Klinken, Guo, Buettner,
  Haider, Bian, Sun, Rosen, and Zaidi]{liu_blocking_2017}
Peng Liu, Yaoting Ji, Tony Yuen, Elizabeth Rendina-Ruedy, Victoria~E.
  {DeMambro}, Samarth Dhawan, Wahid Abu-Amer, Sudeh Izadmehr, Bin Zhou,
  Andrew~C. Shin, Rauf Latif, Priyanthan Thangeswaran, Animesh Gupta, Jianhua
  Li, Valeria Shnayder, Samuel~T. Robinson, Yue~Eric Yu, Xingjian Zhang, Feiran
  Yang, Ping Lu, Yu~Zhou, Ling-Ling Zhu, Douglas~J. Oberlin, Terry~F. Davies,
  Michaela~R. Reagan, Aaron Brown, T.~Rajendra Kumar, Solomon Epstein, Jameel
  Iqbal, Narayan~G. Avadhani, Maria~I. New, Henrik Molina, Jan~B. van Klinken,
  Edward~X. Guo, Christoph Buettner, Shozeb Haider, Zhuan Bian, Li~Sun,
  Clifford~J. Rosen, and Mone Zaidi.
\newblock Blocking {FSH} induces thermogenic adipose tissue and reduces body
  fat.
\newblock \emph{Nature}, 546:\penalty0 107--112, 2017.
\newblock ISSN 1476-4687.
\newblock \doi{10.1038/nature22342}.

\bibitem[Henderson et~al.(2000)Henderson, Diggle, and
  Dobson]{henderson_joint_2000}
R.~Henderson, P.~Diggle, and A.~Dobson.
\newblock Joint modelling of longitudinal measurements and event time data.
\newblock \emph{Biostatistics}, 1:\penalty0 465--480, 2000.
\newblock ISSN 1465-4644.
\newblock \doi{10.1093/biostatistics/1.4.465}.

\bibitem[Proust-Lima et~al.(2014)Proust-Lima, Séne, Taylor, and
  Jacqmin-Gadda]{proust-lima_joint_2014}
Cécile Proust-Lima, Mbéry Séne, Jeremy M.~G. Taylor, and Hélène
  Jacqmin-Gadda.
\newblock Joint latent class models for longitudinal and time-to-event data: a
  review.
\newblock \emph{Statistical Methods in Medical Research}, 23:\penalty0 74--90,
  2014.
\newblock ISSN 1477-0334.
\newblock \doi{10.1177/0962280212445839}.

\bibitem[Wang et~al.(2017)Wang, Luo, and Li]{wang_dynamic_2017}
Jue Wang, Sheng Luo, and Liang Li.
\newblock Dynamic prediction for multiple repeated measures and event time
  data: An application to parkinson’s disease.
\newblock \emph{The Annals of Applied Statistics}, 11:\penalty0 1787--1809,
  2017.
\newblock ISSN 1932-6157, 1941-7330.
\newblock \doi{10.1214/17-AOAS1059}.
\newblock URL
  \url{https://projecteuclid.org/journals/annals-of-applied-statistics/volume-11/issue-3/Dynamic-prediction-for-multiple-repeated-measures-and-event-time-data/10.1214/17-AOAS1059.full}.

\bibitem[Carroll(2003)]{carroll_variances_2003}
Raymond Carroll.
\newblock Variances are not always nuisance parameters.
\newblock \emph{Biometrics}, 59:\penalty0 211--220, 2003.
\newblock \doi{10.1111/1541-0420.t01-1-00027}.

\bibitem[Young and Benton(2018)]{young_heart-rate_2018}
Hayley~A. Young and David Benton.
\newblock Heart-rate variability: a biomarker to study the influence of
  nutrition on physiological and psychological health?
\newblock \emph{Behavioural Pharmacology}, 29\penalty0 (2):\penalty0 140--151,
  2018.
\newblock ISSN 0955-8810.
\newblock \doi{10.1097/FBP.0000000000000383}.
\newblock URL \url{https://www.ncbi.nlm.nih.gov/pmc/articles/PMC5882295/}.

\bibitem[Ogburn et~al.(2021)Ogburn, Rudolph, Morello-Frosch, Khan, and
  Casey]{ogburn2021warning}
Elizabeth~L Ogburn, Kara~E Rudolph, Rachel Morello-Frosch, Amber Khan, and
  Joan~A Casey.
\newblock A warning about using predicted values from regression models for
  epidemiologic inquiry.
\newblock \emph{American Journal of Epidemiology}, 190:\penalty0 1142--1147,
  2021.
\newblock \doi{10.1093/aje/kwaa282}.

\bibitem[Wang et~al.(2020)Wang, McCormick, and Leek]{wang2020methods}
Siruo Wang, Tyler~H McCormick, and Jeffrey~T Leek.
\newblock Methods for correcting inference based on outcomes predicted by
  machine learning.
\newblock \emph{Proceedings of the National Academy of Sciences}, 117:\penalty0
  30266--30275, 2020.

\bibitem[Chi and Ibrahim(2006)]{chi_joint_2006}
Yueh-Yun Chi and Joseph~G. Ibrahim.
\newblock Joint models for multivariate longitudinal and multivariate survival
  data.
\newblock \emph{Biometrics}, 62:\penalty0 432--445, 2006.
\newblock ISSN 0006-341X.
\newblock \doi{10.1111/j.1541-0420.2005.00448.x}.

\bibitem[Ibrahim et~al.(2010)Ibrahim, Chu, and Chen]{ibrahim_basic_2010}
Joseph~G. Ibrahim, Haitao Chu, and Liddy~M. Chen.
\newblock Basic concepts and methods for joint models of longitudinal and
  survival data.
\newblock \emph{Journal of Clinical Oncology}, 28:\penalty0 2796--2801, 2010.
\newblock ISSN 0732-183X.
\newblock \doi{10.1200/JCO.2009.25.0654}.
\newblock URL \url{https://www.ncbi.nlm.nih.gov/pmc/articles/PMC4503792/}.

\bibitem[Lawrence~Gould et~al.(2015)Lawrence~Gould, Boye, Crowther, Ibrahim,
  Quartey, Micallef, and Bois]{lawrence_gould_joint_2015}
A.~Lawrence~Gould, Mark~Ernest Boye, Michael~J. Crowther, Joseph~G. Ibrahim,
  George Quartey, Sandrine Micallef, and Frederic~Y. Bois.
\newblock Joint modeling of survival and longitudinal non-survival data:
  current methods and issues; {R}eport of the {DIA} {B}ayesian joint modeling
  working group.
\newblock \emph{Statistics in Medicine}, 34:\penalty0 2181--2195, 2015.
\newblock ISSN 1097-0258.
\newblock \doi{10.1002/sim.6141}.

\bibitem[Long and Mills(2018)]{long_joint_2018}
Jeffrey~D. Long and James~A. Mills.
\newblock Joint modeling of multivariate longitudinal data and survival data in
  several observational studies of huntington’s disease.
\newblock \emph{{BMC} Medical Research Methodology}, 18:\penalty0 138, 2018.
\newblock ISSN 1471-2288.
\newblock \doi{10.1186/s12874-018-0592-9}.
\newblock URL \url{https://doi.org/10.1186/s12874-018-0592-9}.

\bibitem[Papageorgiou et~al.(2019)Papageorgiou, Mauff, Tomer, and
  Rizopoulos]{Papageorgiou2019}
Grigorios Papageorgiou, Katya Mauff, Anirudh Tomer, and Dimitris Rizopoulos.
\newblock An overview of joint modeling of time-to-event and longitudinal
  outcomes.
\newblock \emph{Annual Review of Statistics and Its Application}, 6:\penalty0
  223--240, 2019.
\newblock \doi{10.1146/annurev-statistics-030718-105048}.
\newblock URL \url{https://doi.org/10.1146/annurev-statistics-030718-105048}.

\bibitem[Elliott et~al.(2012)Elliott, Sammel, and
  Faul]{elliott_associations_2012}
Michael~R. Elliott, Mary~D. Sammel, and Jessica Faul.
\newblock Associations between variability of risk factors and health outcomes
  in longitudinal studies.
\newblock \emph{Statistics in Medicine}, 31:\penalty0 2745--2756, 2012.
\newblock ISSN 0277-6715.
\newblock \doi{10.1002/sim.5370}.
\newblock URL \url{https://www.ncbi.nlm.nih.gov/pmc/articles/PMC3470883/}.

\bibitem[Harlow et~al.(2000)Harlow, Lin, and Ho]{harlow_analysis_2000}
S.~D. Harlow, X.~Lin, and M.~J. Ho.
\newblock Analysis of menstrual diary data across the reproductive life span
  applicability of the bipartite model approach and the importance of
  within-woman variance.
\newblock \emph{Journal of Clinical Epidemiology}, 53:\penalty0 722--733, 2000.
\newblock ISSN 0895-4356.
\newblock \doi{10.1016/s0895-4356(99)00202-4}.

\bibitem[Huang et~al.(2014)Huang, Elliott, and Harlow]{huang_modeling_2014}
Xiaobi Huang, Michael~R. Elliott, and Siobán~D. Harlow.
\newblock Modeling menstrual cycle length and variability at the approach of
  menopause using hierarchical change point models.
\newblock \emph{Journal of the Royal Statistical Society. Series C, Applied
  Statistics}, 63:\penalty0 445--466, 2014.
\newblock ISSN 0035-9254.
\newblock \doi{10.1111/rssc.12044}.

\bibitem[Sammel et~al.(2001)Sammel, Wang, Ratcliie, Freeman, and
  Propert]{sammel_models_2001}
Mary Sammel, Y~Wang, S~Ratcliie, Ellen Freeman, and K~Propert.
\newblock Models for within-subject heterogeneity as predictors for disease.
\newblock In \emph{Proceedings of the Annual Meeting of the American
  Statistical Association}, 2001.

\bibitem[Jiang et~al.(2015)Jiang, Elliott, Sammel, and Wang]{jiang_joint_2015}
Bei Jiang, Michael~R. Elliott, Mary~D. Sammel, and Naisyin Wang.
\newblock Joint modeling of cross-sectional health outcomes and longitudinal
  predictors via mixtures of means and variances.
\newblock \emph{Biometrics}, 71:\penalty0 487--497, 2015.
\newblock ISSN 1541-0420.
\newblock \doi{https://doi.org/10.1111/biom.12284}.
\newblock URL \url{http://onlinelibrary.wiley.com/doi/abs/10.1111/biom.12284}.

\bibitem[Carroll et~al.(2006)Carroll, Ruppert, Stefanski, and
  Crainiceanu]{carroll_measurement_2006}
Raymond~J. Carroll, David Ruppert, Leonard~A. Stefanski, and Ciprian~M.
  Crainiceanu.
\newblock \emph{Measurement Error in Nonlinear Models: A Modern Perspective,
  Second Edition}.
\newblock {CRC} Press {LLC}, 2006.
\newblock ISBN 978-1-4200-1013-8.
\newblock URL
  \url{http://ebookcentral.proquest.com/lib/umichigan/detail.action?docID=274076}.

\bibitem[Griliches and Intriligator(1987)]{griliches_chapter_1986}
Zvi Griliches and Michael~D. Intriligator.
\newblock \emph{Handbook of Econometrics}, chapter~25, pages 1465--1514.
\newblock North Holland, 1987.

\bibitem[Gelman(2006)]{gelman_prior_2006}
Andrew Gelman.
\newblock Prior distributions for variance parameters in hierarchical models
  (comment on article by {B}rowne and {D}raper).
\newblock \emph{Bayesian Analysis}, 1:\penalty0 515--534, 2006.
\newblock ISSN 1936-0975, 1931-6690.
\newblock \doi{10.1214/06-BA117A}.
\newblock URL
  \url{http://projecteuclid.org/journals/bayesian-analysis/volume-1/issue-3/Prior-distributions-for-variance-parameters-in-hierarchical-models-comment-on/10.1214/06-BA117A.full}.

\bibitem[Lewandowski et~al.(2009)Lewandowski, Kurowicka, and
  Joe]{lewandowski2009generating}
Daniel Lewandowski, Dorota Kurowicka, and Harry Joe.
\newblock Generating random correlation matrices based on vines and extended
  onion method.
\newblock \emph{Journal of Multivariate Analysis}, 100:\penalty0 1989--2001,
  2009.
\newblock \doi{https://doi.org/10.1016/j.jmva.2009.04.008}.

\bibitem[Ghosh et~al.(2021)Ghosh, Mallick, and Pourahmadi]{ghosh_bayesian_2021}
Riddhi~Pratim Ghosh, Bani Mallick, and Mohsen Pourahmadi.
\newblock Bayesian estimation of correlation matrices of longitudinal data.
\newblock \emph{Bayesian Analysis}, 16:\penalty0 1039--1058, 2021.
\newblock ISSN 1936-0975, 1931-6690.
\newblock \doi{10.1214/20-BA1237}.
\newblock URL
  \url{https://projecteuclid.org/journals/bayesian-analysis/volume-16/issue-3/Bayesian-Estimation-of-Correlation-Matrices-of-Longitudinal-Data/10.1214/20-BA1237.full}.

\bibitem[Carpenter et~al.(2017)Carpenter, Gelman, Hoffman, Lee, Goodrich,
  Betancourt, Brubaker, Guo, Li, and Riddell]{carpenter_stan_2017}
Bob Carpenter, Andrew Gelman, Matthew~D. Hoffman, Daniel Lee, Ben Goodrich,
  Michael Betancourt, Marcus Brubaker, Jiqiang Guo, Peter Li, and Allen
  Riddell.
\newblock Stan: A probabilistic programming language.
\newblock \emph{Journal of Statistical Software}, 76:\penalty0 1--32, 2017.
\newblock \doi{10.18637/jss.v076.i01}.
\newblock URL
  \url{https://www.jstatsoft.org/index.php/jss/article/view/v076i01}.

\bibitem[Chen et~al.(2022)Chen, Wu, Harlow, Karvonen-Gutierrez, Hood, and
  Elliott]{chen_supp_2022}
Irena Chen, Zhenke Wu, Siobán~D. Harlow, Carrie~A. Karvonen-Gutierrez,
  Michelle~M. Hood, and Michael~R. Elliott.
\newblock Supplement to variance as a predictor of health outcomes: using
  subject-level trajectories and variability of sex hormones to predict body
  fat changes in peri- and post-menopausal women.
\newblock 2022.

\bibitem[{Stan Development Team}(2020)]{rstan}
{Stan Development Team}.
\newblock {RStan}: the {R} interface to {Stan}, 2020.
\newblock URL \url{http://mc-stan.org/}.
\newblock R package version 2.21.2.

\bibitem[Vehtari et~al.(2021)Vehtari, Gelman, Simpson, Carpenter, and
  Bürkner]{vehtari_rank-normalization_2021}
Aki Vehtari, Andrew Gelman, Daniel Simpson, Bob Carpenter, and Paul-Christian
  Bürkner.
\newblock Rank-normalization, folding, and localization: An improved rˆ for
  assessing convergence of {MCMC} (with discussion).
\newblock \emph{Bayesian Analysis}, 16:\penalty0 667--718, 2021.
\newblock ISSN 1936-0975, 1931-6690.
\newblock \doi{10.1214/20-BA1221}.
\newblock URL
  \url{https://projecteuclid.org/journals/bayesian-analysis/volume-16/issue-2/Rank-Normalization-Folding-and-Localization--An-Improved-R%cb%86-for/10.1214/20-BA1221.full}.
\newblock Publisher: International Society for Bayesian Analysis.

\bibitem[Gelman et~al.(2013)Gelman, Carlin, Stern, Dunson, Vehtari, and
  Rubin]{gelman_bayesian_2013}
Andrew Gelman, John~B. Carlin, Hal~S. Stern, David~B. Dunson, Aki Vehtari, and
  Donald~B. Rubin.
\newblock \emph{Bayesian Data Analysis}.
\newblock {CRC} Press, 2013.
\newblock ISBN 978-1-4398-9820-8.

\bibitem[Bürkner(2017)]{brms2017}
Paul-Christian Bürkner.
\newblock {brms}: An {R} package for {Bayesian} multilevel models using {Stan}.
\newblock \emph{Journal of Statistical Software}, 80:\penalty0 1--28, 2017.
\newblock \doi{10.18637/jss.v080.i01}.

\bibitem[Sowers et~al.(2000)Sowers, Crawford, Sternfeld, Morganstein, Gold,
  Greendale, Evans, Neer, Matthews, Sherman, Lo, Weiss, and
  Kelsey]{sowers_chapter_2000}
{MARYFRAN} Sowers, {SYBIL}~L. Crawford, {BARBARA} Sternfeld, {DAVID}
  Morganstein, {ELLEN}~B. Gold, {GAIL}~A. Greendale, {DENIS} Evans, {ROBERT}
  Neer, {KAREN} Matthews, {SHERRY} Sherman, {ANNIE} Lo, {GERSON} Weiss, and
  {JENNIFER} Kelsey.
\newblock Swan: A multicenter, multiethnic, community-based cohort study of
  women and the menopausal transition.
\newblock In {ROGERIO}~A. Lobo, {JENNIFER} Kelsey, and {ROBERT} Marcus,
  editors, \emph{Menopause: Biology and Pathology}, pages 175--188. Academic
  Press, 2000.
\newblock ISBN 978-0-12-453790-3.
\newblock \doi{10.1016/B978-012453790-3/50012-3}.
\newblock URL
  \url{https://www.sciencedirect.com/science/article/pii/B9780124537903500123}.

\bibitem[Greendale et~al.(2019)Greendale, Sternfeld, Huang, Han,
  Karvonen-Gutierrez, Ruppert, Cauley, Finkelstein, Jiang, and
  Karlamangla]{greendale_changes_2019}
Gail~A. Greendale, Barbara Sternfeld, {MeiHua} Huang, Weijuan Han, Carrie
  Karvonen-Gutierrez, Kristine Ruppert, Jane~A. Cauley, Joel~S. Finkelstein,
  Sheng-Fang Jiang, and Arun~S. Karlamangla.
\newblock Changes in body composition and weight during the menopause
  transition.
\newblock \emph{{JCI} Insight}, 4:\penalty0 e124865, 2019.
\newblock ISSN 2379-3708.
\newblock \doi{10.1172/jci.insight.124865}.
\newblock URL \url{https://insight.jci.org/articles/view/124865}.

\bibitem[Greendale et~al.(2021)Greendale, Han, Finkelstein, Burnett-Bowie,
  Huang, Martin, and Karlamangla]{greendale_changes_2021}
Gail~A. Greendale, Weijuan Han, Joel~S. Finkelstein, Sherri-Ann~M.
  Burnett-Bowie, {MeiHua} Huang, Deborah Martin, and Arun~S. Karlamangla.
\newblock Changes in regional fat distribution and anthropometric measures
  across the menopause transition.
\newblock \emph{The Journal of Clinical Endocrinology and Metabolism},
  106:\penalty0 2520--2534, 2021.
\newblock ISSN 1945-7197.
\newblock \doi{10.1210/clinem/dgab389}.

\bibitem[Pettee~Gabriel et~al.(2017)Pettee~Gabriel, Sternfeld, Colvin, Stewart,
  Strotmeyer, Cauley, Dugan, and
  Karvonen-Gutierrez]{pettee_gabriel_physical_2017}
Kelley Pettee~Gabriel, Barbara Sternfeld, Alicia Colvin, Andrea Stewart,
  Elsa~S. Strotmeyer, Jane~A. Cauley, Sheila Dugan, and Carrie
  Karvonen-Gutierrez.
\newblock Physical activity trajectories during midlife and subsequent risk of
  physical functioning decline in late mid-life: The {S}tudy of {W}omen's
  {H}ealth {A}cross the {N}ation ({SWAN}).
\newblock \emph{Preventive Medicine}, 105:\penalty0 287--294, 2017.
\newblock ISSN 1096-0260.
\newblock \doi{10.1016/j.ypmed.2017.10.005}.

\end{thebibliography}






\end{document}


\setstcolor{red}

\def\spacingset#1{\renewcommand{\baselinestretch}%
{#1}\small\normalsize} \spacingset{1}


  \bigskip
  \bigskip
  \begin{center}

\title{Supplementary Materials for ``Variance as a predictor of health outcomes''} 
\end{center}
 \medskip
\bigskip

\spacingset{1.9} 

\tableofcontents

\section{Visualization of the Joint Model}
\begin{figure}[!htb]
    \centering
    \includegraphics[scale=0.3]{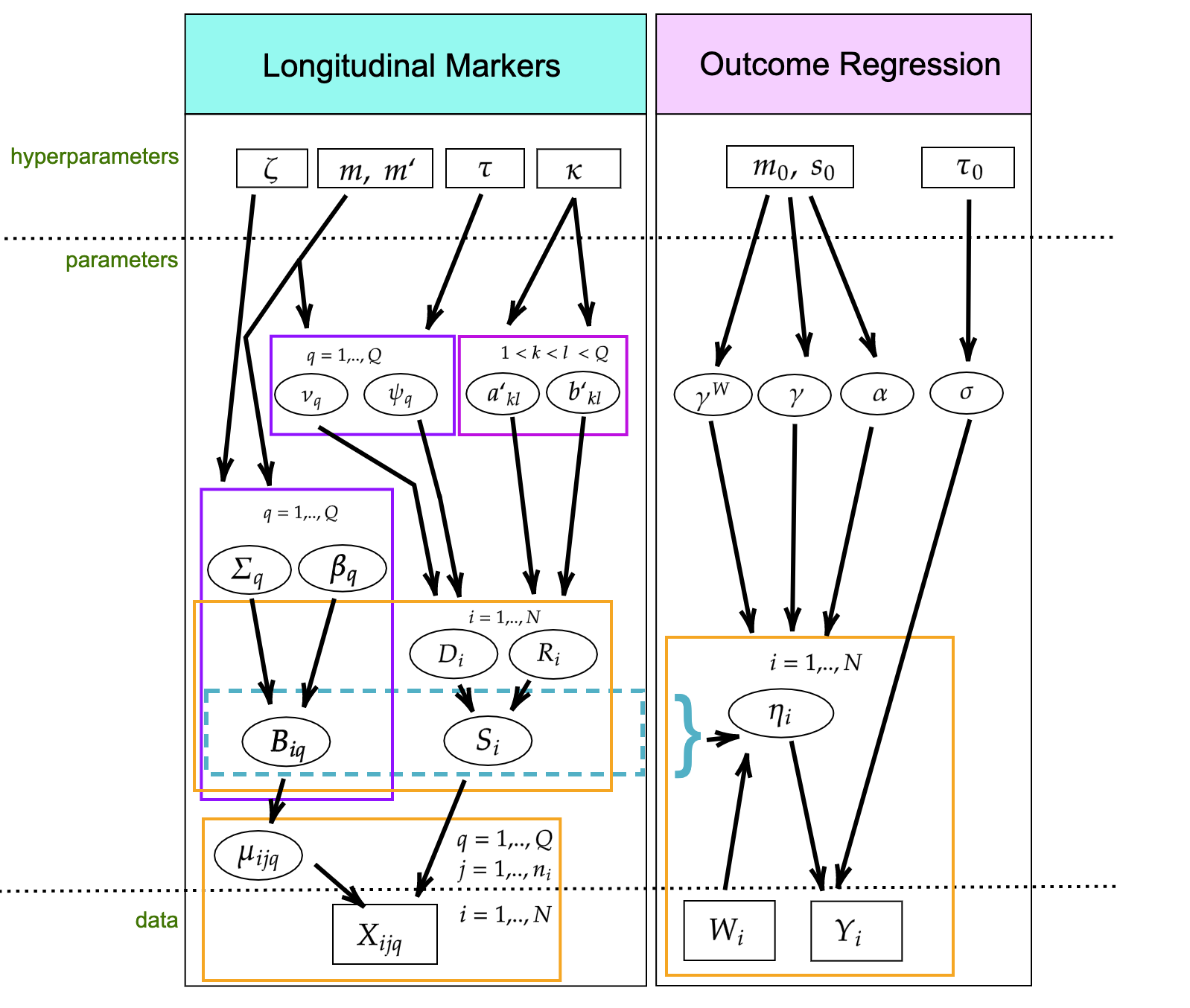}
\caption{A visualization of the relationships between the model parameters and data. This directed graph shows the hierarchical form of our model framework. The quantities in squares are either data or hyperparameters; the unknown quantities are displayed in circles. The arrows connecting variables indicate that the parent parameterizes the distribution of the child node. The rectangular ''plates" that enclose variables indicate that a similar graphical structure is repeated over the index. The index in a plate indicates nodes, hyperparameter levels and subjects.}
    \label{fig:joint_model_dag}
\end{figure}
\newpage
\section{Posterior Predictive Model Checking}
\label{sec::ppc}

To assess our model's validity on the SWAN data, we conduct posterior predictive checks for both the trajectories submodel and the outcome submodel. 

\par For the outcome submodel, we generated simulated data from the posterior predictive distribution. The posterior predictive distribution for the predicted outcome, $\Tilde{Y}$ can be written as:
\begin{align*}
    p(\Tilde{Y}|Y) = \int p(\Tilde{Y}|\theta, X)p(\theta, X|Y)d\theta dX
\end{align*} where $\theta$ are the unknown model parameters and X are the predictor variables used in the outcome regression. For each draw of the model parameters from the posterior distribution, $p(\theta|Y, X)$, we can draw a vector $\Tilde{Y}$ from the posterior predictive distribution by conditioning on the draw of the model parameters and then simulating from the data model \citep{bayesplot_article_2018}.

\begin{figure*}[t!]
    \centering
    \begin{subfigure}
        \centering
        \includegraphics[scale=0.2]{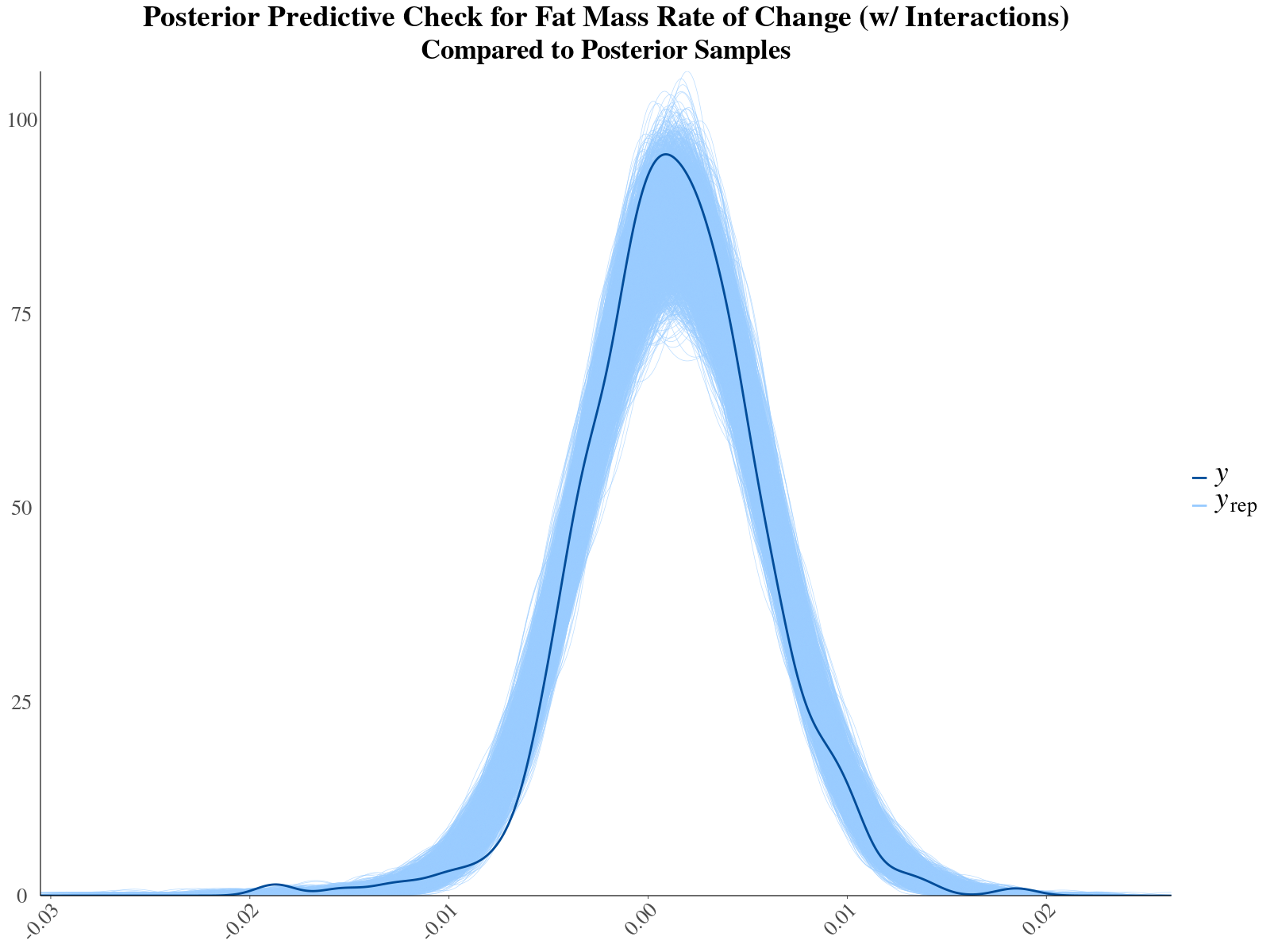}
    \end{subfigure}
    \begin{subfigure}
        \centering
        \includegraphics[scale=0.2]{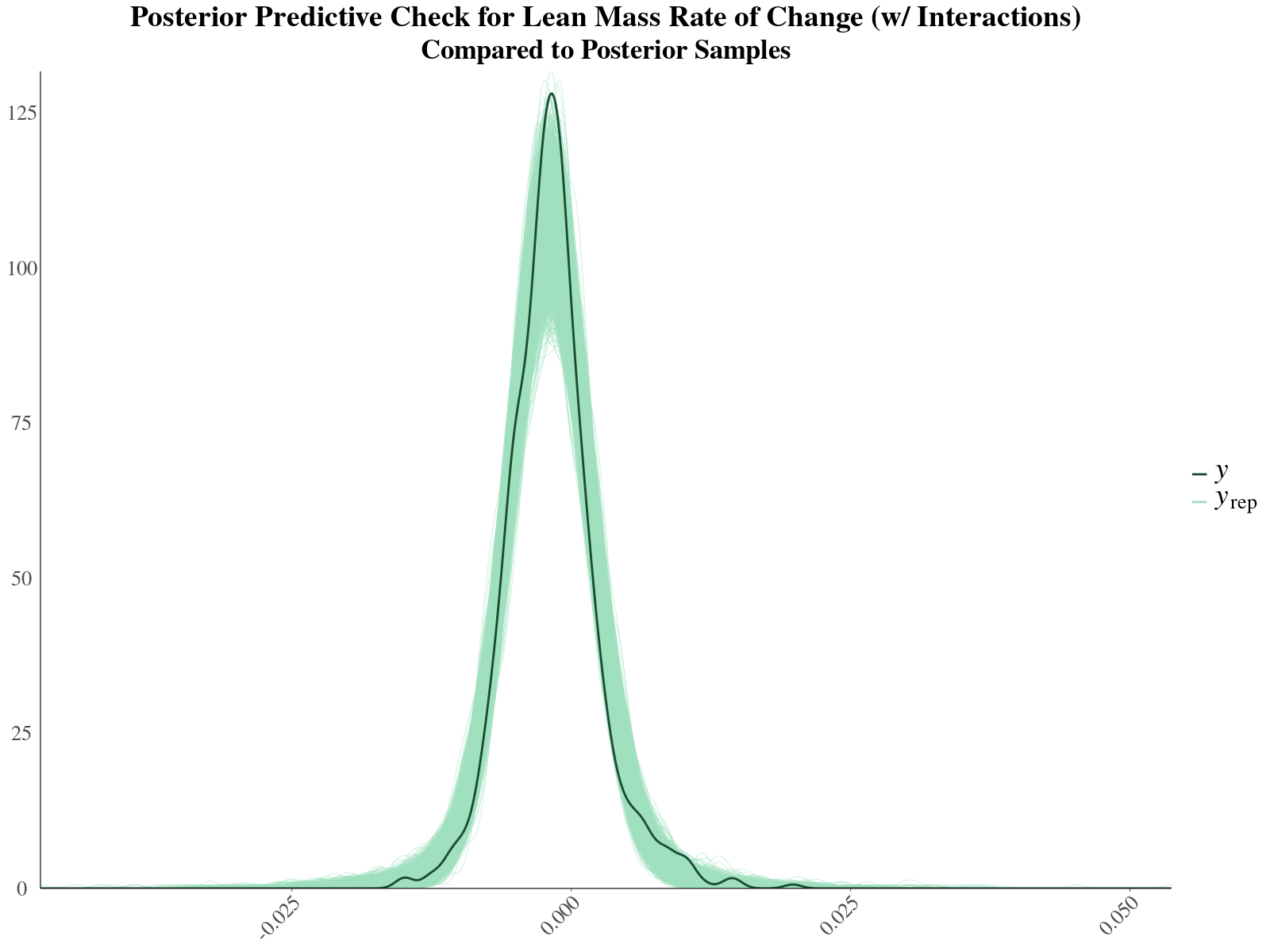}
    \end{subfigure}
    \caption{Visualizations of the posterior predictive checks performed for the fat mass rate of change with variance interactions (top) and lean mass rate of change with variance interactions (bottom). The observed outcomes (y) are represented by the solid lines and the model-generated outcomes (yrep) are represented by the thin semi-opaque lines. We see that the model-generated outcomes cover the observed outcomes for both models, indicating that our model is generating reasonable estimates of the outcomes.}
    \label{fig::ppc_outcome_var_interactions}
\end{figure*}

We then plotted 1,000 draws of this model-generated data against the observed outcome, which is shown in Figure \ref{fig::ppc_outcome_var_interactions}. For both the fat mass rate of change and the lean mass rate of change models, we see that the simulated replicated data from the model overlap the observed data, indicating that our model is producing reasonable predictions. 

\begin{figure}
    \centering
    \includegraphics[scale=0.13]{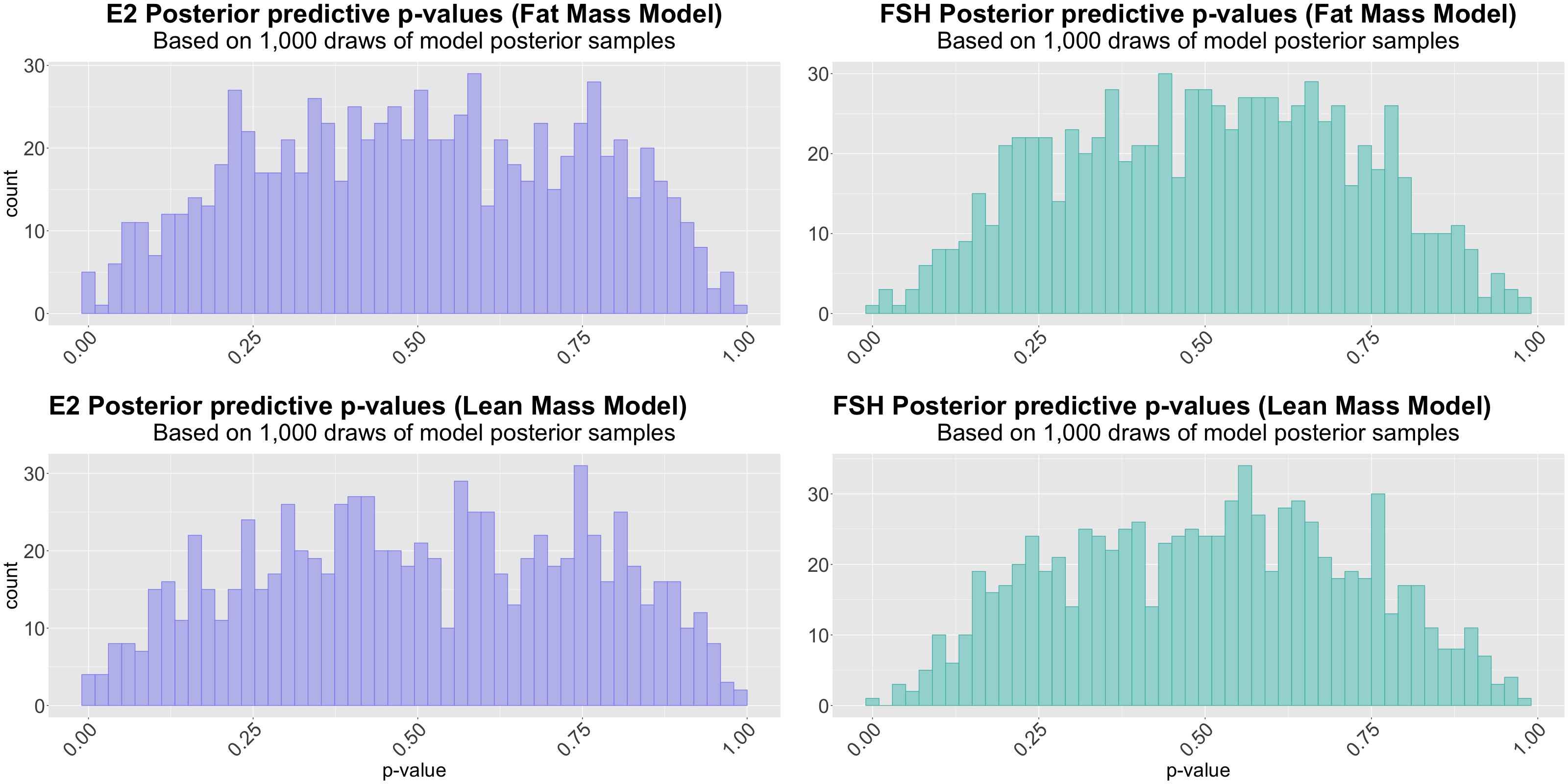}
    \caption{Posterior predictive check of E2, FSH trajectories across all individuals for both the fat mass and lean mass models. The median p-value for each 1,000 draws of posterior samples was 0.5.}
    \label{fig:ppc_hormone_histograms_2x2}
\end{figure}
\par For the trajectories submodel, we define the following statistic:  $T(x_{itq}; b_{iqp}, t)_q = \sum_t (x_{itq} - \mu(b_{iqp},t))^2 /(\sigma_{iq}^2))$ where $\mu(b_{iqp},t))$ is the estimated individual $i$'s mean trajectory for hormone $q$ and  $\sigma^2_{iq}$ is the estimated variance of individual $i$'s trajectory for hormone $q$. By doing this, we can compare $T_i(x^{obs}_{itq}; b_{iqp}, t)_q$ (which is a function of the observed data and the estimated parameters) with $T_i(x^{sim}_{itq}; b_{ipq}, t)_q$ (which is a function of the model generated data  using the model estimated parameters). If there are large discrepancies between  $T_i(x^{obs}_{itq}; b_{iqp}, t, \sigma_{iq}^2)_q$ and $T_i(x^{sim}_{itq}; b_{iqp}, t, \sigma_{iq}^2)_q$, this could indicate poor model fit \citep{gelman_bayesian_2013}. 

\par One way to compare these two $T(x_{itq})$ statistics is to compute the `posterior predictive p-value', which is $P(T_i(x^{obs}_{itq}; b_{iqp}, t, \sigma_{iq}^2)_q < T_i(x^{sim}_{itq}; b_{iqp}, t, \sigma_{iq}^2)_q | (x^{obs}_{itq}))$. For E2 and FSH, we keep $(x^{obs}_{it})$ fixed at the observed values and compute 1,000 values of $T_i(x^{sim}_{itq}; b_{iqp}, t)$ from the posterior of $ b_{iqp}, t, \sigma_{iq}^2$. We then compare these values with 1,000 draws from $T_i(x^{sim}_{itq}; b_{iqp}, t, \sigma_{iq}^2)_q$. Figure \ref{fig:ppc_hormone_histograms_2x2} displays the histograms of the resulting p-values for each individual's hormone trajectory for the two models. Across all of the hormones, most of the the computed p-values were between 0.25 and 0.75. Further analysis of the p-values across the quantiles of the distribution shows that the generated data from the model reasonably captures the individual trends. This provides justification that both the trajectories submodel and the outcome submodel are good fits for the data.


\newpage 
\section{Data Analysis: Posterior Means and 95\% CrIs for Additional Model Parameters}
\label{supp:model_parms}

In this section, we present the estimated posterior means and 95\% credible intervals for the other parameters from the data application (Section 5 in the main text). Table \ref{tab:supp_fat_mass_pars} contains the estimates for the fat mass rate of change model and  Table \ref{tab:supp_lean_mass_pars} contains the estimates for the lean mass rate of change model. 
 
\begin{table}
    \centering
    \renewcommand{\arraystretch}{0.1}
  \begin{tabular}{l r r r r}
\toprule
\multicolumn{1}{l}{Parameter}&\multicolumn{1}{c}{Post. Mean}&\multicolumn{1}{c}{2.5\% CrI}&\multicolumn{1}{c}{97.5\% CrI}\tabularnewline
\midrule
$\beta_{11}$ & -37.37 & -59.01 & -15.79  &\tabularnewline
$\beta_{12}$ & 3.78  &0.78 & 6.84 &  \tabularnewline
$\beta_{21}$ & 26.89 & 4.06 & 48.02 &  \tabularnewline
$\beta_{22}$ & -6.48  &  -9.00 & -4.08 &\tabularnewline
$\Sigma_{1}[1,1]$ & 48.07 & 39.36 & 57.96 & \tabularnewline
$\Sigma_{1}[1,2]$ & -1.50  & -2.43  & 0.55 &\tabularnewline
$\Sigma_{1}[2,2]$ & 0.54  & 0.36 & 0.72 & \tabularnewline
$\Sigma_{2}[1,1]$ & 80.40 & 70.93 & 91.43  & \tabularnewline
$\Sigma_{2}[1,2]$ & 0.13 & -0.63 & 0.88 & \tabularnewline
$\Sigma_{2}[2,2]$ & 0.56 &0.45 & 0.67 & \tabularnewline
$\nu_1$  & -293.86  &-312.52 & -274.63 & \tabularnewline
$\nu_2$ & -701.88 & -725.78 & -677.63 &\tabularnewline
$\xi_1$ & 151.41 & 131.17 & 172.56 & \tabularnewline
$\xi_2$ & 247.34 & 225.05 & 269.14 & \tabularnewline
$\alpha_{1}*$ & 9.00  &  6.72 & 12.02 & \tabularnewline
$\beta_{1}*$ & 33.87 & 24.83 & 45.69 & \tabularnewline
$\sigma^{2}$ & 3.61 & 3.31 & 3.90 & \tabularnewline
\bottomrule
\end{tabular}
    \caption{Evaluation of the posterior means and 95\% CrI estimates for the other parameters in the fat mass rate of change model. All values except for $\alpha_{1}, \beta_{1}$ (indicated with asterisk)  have been multiplied by $10^3$. $\alpha_{1}, \beta_{1}$ have been presented in their original values. 
    } 
    \label{tab:supp_fat_mass_pars}
\end{table}

\begin{table}
    \centering
    \renewcommand{\arraystretch}{0.1}
   \begin{tabular}{l r r r r}
\toprule
\multicolumn{1}{l}{Parameter}&\multicolumn{1}{c}{Post. Mean}&\multicolumn{1}{c}{2.5\% CrI}&\multicolumn{1}{c}{97.5\% CrI}\tabularnewline
\midrule
$\beta_{11}$  & -37.42 & -58.27 & -16.74 &\tabularnewline
$\beta_{12}$ & 3.67  & 0.47 & 6.70 &  \tabularnewline
$\beta_{21}$ &27.09 & 4.91 & 49.10 & \tabularnewline
$\beta_{22}$  & -6.47 & -8.92 &-3.98  &\tabularnewline
$\Sigma_{1}[1,1]$ & 47.92 &  39.14 & 57.88 & \tabularnewline
$\Sigma_{1}[1,2]$ & -1.54  & -2.44 & -0.66 &\tabularnewline
$\Sigma_{1}[2,2]$  & 0.52 & 0.33 & 0.71 & \tabularnewline
$\Sigma_{2}[1,1]$ & 80.55 & 70.92 & 91.29 &\tabularnewline
$\Sigma_{2}[1,2]$ & 0.08 & -0.69 &  0.84 & \tabularnewline
$\Sigma_{2}[2,2]$ & 0.56 &0.44 & 0.68& \tabularnewline
$\nu_1$  & -293.17 & -312.10 & -273.82 & \tabularnewline
$\nu_2$  &-702.11 &-725.35 & -679.66 &\tabularnewline
$\xi_1$  & 141.88 & 119.81 & 161.74 &  \tabularnewline
$\xi_2$  & 241.88 & 220.67 & 262.82 & \tabularnewline
$\alpha_{1}$*  & 9.15  &  6.79 & 12.06 & \tabularnewline
$\beta_{1}$*  & 34.41 & 24.90 &  46.00 & \tabularnewline
$\sigma^{2}_{1}$  & 2.74  &  2.28 & 3.10 & \tabularnewline
$\sigma^{2}_{2}$   & 6.71 & 4.77 &  10.70 & \tabularnewline
$\Pi_{1}$* & 0.86 & 0.66 & 0.97 & \tabularnewline
\bottomrule
\end{tabular}
    \caption{Evaluation of of the posterior means and 95\% CrI estimates for the other model parameters in the lean mass rate of change model. All values except for $\alpha_{1}, \beta_{1}, \Pi_{1}$ (indicated with asterisk)  have been multiplied by $10^3$. $\alpha_{1}, \beta_{1}, \Pi_{1}$ have been presented in their original values.}
    \label{tab:supp_lean_mass_pars}
\end{table}
\clearpage
\newpage

\section{Two Biomarker Simulation Study and Three Biomarker Simulation Study}

In this section, we present the bias, coverage and average interval length for the other JMIV model parameters from running 200 simulation replicates. For each simulation replicate, we ran two chains with 2,000 steps and 1,000 burn in. The data generation parameters are detailed in Section 4.1 and 4.2 of the main text. Table \ref{tab:sim1_other_pars} contains the estimates for the two biomarker simulation study and Table \ref{tab:sim2_other_pars} contains the estimates for the three biomarker simulation study.

\begin{table}
    \centering
    \renewcommand{\arraystretch}{0.3}
  \begin{tabular}{l r r r r r r}
\toprule
\multicolumn{1}{l}{Parameter}&\multicolumn{1}{c}{Truth}&\multicolumn{1}{c}{Average Post. Mean}&\multicolumn{1}{c}{ Bias}&\multicolumn{1}{c}{Coverage \%}&\multicolumn{1}{c}{Average Interval Length}\tabularnewline
\midrule
$\beta_{11}$ & 0&  0.00 & 0.00& 95.5& 0.14&\tabularnewline
$\beta_{12}$ & 2& 2.00 & 0.00& 95.0 &0.13& \tabularnewline
$\beta_{21}$ &2& 2.00 &0.01 &  95.5 & 0.15 & \tabularnewline
$\beta_{22}$ & 1& 1.00 & 0.00& 95.5& 0.09& \tabularnewline
$\Sigma_{1}[1,1]$ & 1 &  1.00 & 0.00& 95.0 & 0.22  \tabularnewline
$\Sigma_{1}[1,2]$ & -0.05 & -0.05&  0.00&  96.0 & 0.14& \tabularnewline
$\Sigma_{1}[2,2]$ & 1.01&  1.00& 0.01 & 94.0 & 0.19& \tabularnewline
$\Sigma_{2}[1,1]$ & 1& 1.00&  0.00&  96.0 & 0.24&\tabularnewline
$\Sigma_{2}[1,2]$ & -0.1 &  -0.10& 0.00& 97.5 & 0.10& \tabularnewline
$\Sigma_{2}[2,2]$ & 0.5& 0.50&  0.00&  95.0 & 0.10&\tabularnewline
$\nu_1$ & 0&  0.00& 0.00& 95.5& 0.06&\tabularnewline
$\nu_2$ & 0.25& 0.25& 0.00& 97.0 &0.04& \tabularnewline
$\xi_1$ & 0.38& 0.37 & 0.00 &  95.0 & 0.05 & \tabularnewline
$\xi_2$ & 0.25& 0.25& 0.00& 97.0& 0.04& \tabularnewline
$a_{1}$  & 1 &  1.01& 0.01& 96.0 & 0.01& \tabularnewline
$b_{1}$  & 5& 5.06&  0.06&  96.0& 1.32& \tabularnewline
\bottomrule
\end{tabular}
\caption{Two Trajectory Simulation Setting: Evaluation of bias, coverage, and 95\% credible interval length across 200 simulation replicates for the $\Bb_i$ and $\Sb_i$ parameters for the JMIV model. Our model achieves $>$ 90\% coverage across all parameters and maintains low bias.}
    \label{tab:sim1_other_pars}
\end{table}

\begin{table}
    \centering
    \renewcommand{\arraystretch}{0.3}
   \begin{tabular}{l r r r r r r}
\toprule
\multicolumn{1}{l}{Parameter}&\multicolumn{1}{c}{Truth}&\multicolumn{1}{c}{Average Post. Mean}&\multicolumn{1}{c}{Bias}&\multicolumn{1}{c}{Coverage \%}&\multicolumn{1}{c}{Average Interval Length}\tabularnewline
\midrule
$\beta_{11}$ & 0&  0.00 & 0.00& 95.0& 0.14&\tabularnewline
$\beta_{12}$ & 2& 1.99& 0.00& 93.5 &0.35& \tabularnewline
$\beta_{21}$ &2& 1.99 & 0.00 &  96.0 & 0.14& \tabularnewline
$\beta_{22}$  & 1& 0.99 & 0.00& 95.5& 0.09& \tabularnewline
$\beta_{31}$ &1& 1.99 & 0.00 &  93.0 & 0.14 & \tabularnewline
$\beta_{32}$ & 1& 1.00 & 0.00& 95.5& 0.13& \tabularnewline
$\Sigma_{1}[1,1]$ & 1 &  1.00 & 0.00& 94.0 & 0.21& \tabularnewline
$\Sigma_{1}[1,2]$ & -0.05 & -0.05&  0.00&  97.0& 0.14&\tabularnewline
$\Sigma_{1}[2,2]$ & 1.00&  0.99 & 0.01& 96.0 & 0.18& \tabularnewline
$\Sigma_{2}[1,1]$ & 1& 1.00& 0.00& 96.5 & 0.22&\tabularnewline
$\Sigma_{2}[1,2]$ & -0.1 & -0.10& 0.00& 93.5 & 0.10& \tabularnewline
$\Sigma_{2}[2,2]$  & 0.5& 0.50&  0.00&  97.5 & 0.09&\tabularnewline
$\Sigma_{3}[1,1]$ & 1& 1.00&  0.00&  96.0 & 0.21&\tabularnewline
$\Sigma_{3}[1,2]$ & -0.25 &  -0.25& 0.00& 94.5 & 0.14& \tabularnewline
$\Sigma_{3}[2,2]$  & 1& 1.00&  0.00&  92.0 & 0.18&\tabularnewline
$\nu_1$ & 0.00&  0.00& 0.00& 94.0& 0.06&\tabularnewline
$\nu_2$ & 0.25& 0.25& 0.00& 95.5 &0.04& \tabularnewline
$\nu_3$ & 0.25& 0.25& 0.00& 94.5 &0.07& \tabularnewline
$\xi_1$ & 0.375& 0.38 & 0.00 & 97.0 & 0.06& \tabularnewline
$\xi_2$ & 0.25& 0.25& 0.00& 94.5 & 0.04& \tabularnewline
$\xi_3$ & 0.25& 0.25& 0.00& 94.5 & 0.07& \tabularnewline
$a_{12}$ & 1 &  1.01& 0.01& 96.0 & 0.19& \tabularnewline
$a_{13}$ & 1 &  1.01& 0.01& 94.5 & 0.19& \tabularnewline
$a_{23}$ & 2 &  2.04& 0.04& 93.5 & 0.50& \tabularnewline
$b_{12}$ & 5& 5.09&  0.09&  96.5 & 1.29& \tabularnewline
$b_{13}$ & 5& 5.07& 0.07&92.5& 1.29& \tabularnewline
$b_{23}$ & 2& 2.05 & 0.05& 94.0 & 0.50& \tabularnewline
\bottomrule
\end{tabular}
\caption{Three Trajectory Simulation Setting: Evaluation of bias, coverage, and 95\% credible interval length across 200 simulation replicates for the $\Bb_i$ and $\Sb_i$ parameters for the JMIV model. Our model achieves $>$ 90\% coverage across all parameters and maintains low bias.}
    \label{tab:sim2_other_pars}
\end{table}

\section{Simulation 3: Linear Approximation of Nonlinearity}
\label{supplement_sim3_linear_app}

In this simulation study, we study how well our model performs when the true relationship between some of the longitudinal means and variances terms and the cross-sectional outcome is nonlinear, but we approximate this relationship with a linear form. 

\subsubsection*{Step 1: Estimate Linear Approximation Coefficients} We use the same data generation parameters for the longitudinal markers as in Section 4.1 of the main text. For the outcome model, we generate the mean $\eta(\bB_i, \bS_i)$ as: 

\begin{align*}
  & \eta(\Bb_i, \Sb_i) = 2b_{i11}  +b_{i12} - b_{i21} + 0.5 b_{i22} + 2s_{i11} - s_{i21} + 2s_{i22} +  0.5 b_{i21}^{2} + 0.75s_{i11}^{2}
\end{align*}
so that the individual slope of the first biomarker ($b_{i12})$ and the variance of the first biomarker ($s_{i11}$) are quadratically related to the outcome. 

To estimate the ``linear approximation'' coefficients, we simulate data for 1 million individuals and generate the outcome data as: 
\begin{align*}
    Y_i \sim \mathcal{N}( \eta(\Bb_i, \Sb_i), 0.01)
\end{align*}
We then fit a linear approximation using the \lstinline{lm()} function in R and the following model: 
\begin{align*}
  Y_{i}  = \alpha_{11}* b_{i11} + \alpha_{12}* b_{i12} + \alpha_{21}* b_{i21} + \alpha_{22}* b_{i22} + \gamma_{11}* s_{i11} + \gamma_{21}* s_{i21} + \gamma_{22}* s_{i22} 
\end{align*}
and collect the estimated coefficients. These coefficients are the ``target'' coefficients that we want to approximate. These targets are shown in Table \ref{tab:sim3_alpha_gamma} as the ``truth" values. 

\subsubsection*{Step 2: Joint Model Simulation Replicates}
We follow the same data generation in Step 1 for the longitudinal markers. For the outcome model, we generate the outcome data as: 
\begin{align*}
    Y_i \sim \mathcal{N}( \eta(\Bb_i, \Sb_i), 0.5)
\end{align*}

After generating this data, we apply our joint model, but modeled with the same linear mean function, $\eta(\Bb_i, \Sb_i)$, as in Section 4 of main text and collect the estimated coefficients. We do this for 200 independent replicates and evaluate model performance using the same criteria (bias, coverage, and average 95\% CrI length) as in the previous simulation studies. Table \ref{tab:sim3_alpha_gamma}  displays the results for the outcome mean and variance parameters. We find that the model maintains low bias and high coverage of the truth (> 90\% coverage). This indicates that our model can recover the estimated parameters from a linear approximation of the model, when the true form of the outcome mean may be nonlinear.  
\begin{table}[!ht]
\centering
\renewcommand{\arraystretch}{0.5}
\begin{tabular}{r r r r}
\toprule
Truth &{Bias}&{Coverage (\%)}&{Average Interval Length}\\
\midrule
$\alpha_{11}$ = 1.99&0.02&93.50&0.41\\
\addlinespace
$\alpha_{12}$ = 1.44&0.03&93.00&0.34\\
\addlinespace
$\alpha_{21}$ = -1.06&0.00&94.58&0.38\\
\addlinespace
$\alpha_{22}$ = 0.56&-0.03&94.50&0.51\\
\addlinespace
$\gamma_{11}$ = 1.97&-0.01&96.00&0.69\\
\addlinespace
$\gamma_{12}$ = -0.99&0.03&94.50&1.17\\
\addlinespace
$\gamma_{22}$ = -2.06&-0.03&92.50&0.58\\
\bottomrule
\end{tabular}
\caption{Simulation III: bias, coverage, and 95\% credible interval (or confidence interval) length across 200 simulation replicates. With the linear approximation, our model maintains low bias and high coverage of the true (linear approximating) parameters.}
 \label{tab:sim3_alpha_gamma}
\end{table}

\FloatBarrier
\bibliography{references}